\def\fm{{\hbox{fm}}}

\input epsf

\font\bigboldiii=cmbx10 scaled\magstep3
\font\smalli=cmr8 scaled\magstep1
\font\smallii=cmr8 scaled\magstep2
\font\smallitii=cmmi7 scaled\magstep2

\newcount\fignum

\def\noblackbox{\overfullrule=0pt}

\def\rr{{\bf r}}
\def\BB{{\bf B}}
\def\EE{{\bf E}}
\def\acal{{\cal A}}
\def\bcal{{\cal B}}
\def\dcal{{\cal D}}
\def\ecal{{\cal E}}
\def\lcal{{\cal L}}
\def\ncal{{\cal N}}
\def\pcal{{\cal P}}
\def\qcal{{\cal Q}}
\def\scal{{\cal S}}
\def\tcal{{\cal T}}
\def\vcal{{\cal V}}

\def\and{{\it\&}}
\def\mev{~\hbox{MeV}}
\def\gev{~\hbox{GeV}}
\def\im{{\bf Im}}
\def\lesim{\,{\raise-3pt\hbox{$\sim$}}\!\!\!\!\!{\raise2pt\hbox{$<$}}\,}
\def\lhs{left hand side\ }
\def\quarter{{1\over4}}
\def\rBB{ \hbox{{\smallii I}}\!\hbox{{\smallii R}} }

\def\su#1{{SU(#1)}}
\def\inv#1{{1\over#1}}
\def\deriva#1#2#3{\left({\partial #1\over\partial #2}\right)_{#3}}
\def\lowti#1{_{{\rm #1 }}}
\def\thecaphyz#1{\global\advance\fignum by 1
\centerline{\hfil\vbox to 1 in{\hsize 5 in \vfil
{$\mskip -52.24mu$\textindent{\smallitii Fig:{\rm\ }
\the\fignum{\rm\ }}
        \global \advance \baselineskip by -10 pt
        \tenpoint\noindent #1 }}\hfil}\global \advance \baselineskip by 10 pt
\bigskip}

\def\thetitle{\bigskip\bigskip\bigskip The Thomas-Fermi approximation for
gauge theories}
\def\theabstract{
An effective field approximation, similar to the atomic Thomas-Fermi
approach, is proposed for studying non-Abelian gauge theories which
includes finite-volume effects. As applications of the
formalism the equation of state for an
$\su2$ gauge theory with massless fermions is obtained.
The extensions to realistic situations are briefly discussed.}
\def\ucrnum{173}
\def\theauthor{\advance \baselineskip by -7 pt
\author{{\fourteencp D.D. Dixon\foot{dixon@agouti.ucr.edu}}
\address{{\it Inst. of Geophysics \and\ Planetary Physics \break
University of California at Riverside\break
Riverside, California 92521; U.S.A.}}
\andauthor{{\fourteencp P. Kaus\foot{pkaus@bslnet.com} and
J. Wudka\foot{jose.wudka@ucr.edu}}}}
\address{{\it Physics Department\break
University of California, Riverside\break
Riverside, California 92521--0413; U.S.A.}
\break\medskip\noindent $\mskip-250mu$ {\smalli PACS: 12.38.Mh, 11.10.Wx,
11.15.Tk}}
\advance \baselineskip by 7 pt}

\newlinechar=`\^^J
\immediate\write16{^^J ONE COLUMN OUTPUT  ^^J}

\input phyzzx
\Twelvepoint
\PHYSREV

\rightline{UCRHEP-T\ucrnum}
{\titlepage
\vskip -.2 in
\title{ {\bigboldiii \thetitle}}
\doublespace
\author{\theauthor}
\abstract
\bigskip
\singlespace
\theabstract
\endpage}  \doublespace

\def\whatjournal{P}

\newlinechar=`\^^J

\def\ordernpb#1#2#3{{\bf#1} (#3) #2}
\if P\whatjournal {\global\def\order#1#2#3{\orderprd{#1}{#2}{#3}}}
                    \immediate\write16{^^J PRD references ^^J}\else 
                   {\global\def\order#1#2#3{\ordernpb{#1}{#2}{#3}}}
                    \immediate\write16{^^J NPB references ^^J} 
\fi

\def\ap#1#2#3{{\rm Ann. Phys.\ }\order{#1}{#2}{#3}}

\def\app#1#2#3{{\rm Acta Phys. Pol. {\bf B}}\order{#1}{#2}{#3}}

\def\jmp#1#2#3{{\rm J. Math. Phys. }\order{#1}{#2}{#3}}

\def\nci#1#2#3{{\rm Nuov. Cim }\order{#1}{#2}{#3}}

\def\npa#1#2#3{{\rm Nucl. Phys. {\bf A}}\order{#1}{#2}{#3}}
\def\npb#1#2#3{{\rm Nucl. Phys. {\bf B}}\order{#1}{#2}{#3}}
\def\pcps#1#2#3{{\rm Proc. Cambridge Phil. Soc.}\order{#1}{#2}{#3}}

\def\plb#1#2#3{{\rm Phys. Lett. {\bf B}}\order{#1}{#2}{#3}}

\def\pnas#1#2#3{{\rm Proc. Nat. Acad. Sci.\ }\order{#1}{#2}{#3}}

\def\pr#1#2#3{{\rm Phys. Rev.\ }\order{#1}{#2}{#3}}

\def\prl#1#2#3{{\rm Phys. Rev. Lett.\ }\order{#1}{#2}{#3}}
\def\pra#1#2#3{{\rm Phys. Rev. {\bf A}}\order{#1}{#2}{#3}}
\def\prd#1#2#3{{\rm Phys. Rev. {\bf D}}\order{#1}{#2}{#3}}

\def\rmp#1#2#3{{\rm Rev. Mod. Phys.\ }\order{#1}{#2}{#3}}
\def\rpp#1#2#3{{\rm Rep. Progr. Phys.\ }\order{#1}{#2}{#3}}
\def\zphysold#1#2#3{{\rm Z. Phys. }\order{#1}{#2}{#3}}

\REF\intro{
D. Gross \etal, \rmp{53}{43}{1981}
L. McLerran, \rmp{58}{1021}{1986}
J.I. Kapusta, {\sl Finite - temperature field theory}; Cambridge
Monographs on Mathematical Physics. Cambridge Univ. Press, (Cambridge, 1989);
.}
\REF\IRbehaviour{A. Linde, \rpp{42}{389}{1979}. See also Refs. \intro.}
\REF\htl{E. Braaten and R.D. Pisarski, \prl{64}{1338}{1990},
\npb{337}{569}{1990}}
\REF\kapusta{See, for example, J.I. Kapusta, Ref. \intro}
\REF\soft{For a recent overview see B. M\"uller, in {\sl 14th International
Conference on Particles and Nuclei (PANIC 96)}, Williamsburg, VA, 22-28 May
1996.}
\REF\mc{For recent reviews see, for example,
F. Karsch, lectures given at {\sl International School of
Physics, `Enrico Fermi', Course 80: Selected Topics in Nonperturbative QCD},
Varenna, Italy, 27 Jun - 7 Jul 1995. (e-Print Archive: hep-lat/9512029).
A. Ukawa (Tsukuba U.), in {\sl Uehling Summer School on
Phenomenology and Lattice QCD}, Seattle, WA, 21 Jun - 2 Jul 1993.
(e-Print Archive: hep-lat/9505024).}
\REF\atomictf{
L.H. Thomas, \pcps{23}{542}{1927}.
E. Fermi, \zphysold{48}{73}{1928}.
P.A.M. Dirac, \pcps{26}{376}{1930}. See also
R.D. Cowan and J. Ashkin, \pr{105}{144}{1957}.
J. Schwinger, \pra{22}{1827}{1980}; \pra{24}{2353}{1981}.
}
\REF\bj{H.A. Bethe and R.W. Jackiw, {\sl Intermediate Quantum Mechanics}
(Benjamin/Cummings, 1986).}
\REF\bag{
A. Chodos \etal, \prd{9}{3471}{1974}; \prd{10}{2599}{1974}.
R.L. Jaffe, Lectures presented at the {\sl  Erice School of Subnuclear
Physics}, Erice, Sicily, Jul 31 - Aug 10, 1979.
J. Goldstone and R.L. Jaffe, \prl{51}{1518}{1983}.
A. Wirzba, Presented at  {\sl 1992 International Nuclear Physics Conference},
Wiesbaden, Germany, 26 Jul - 1 Aug 1992. \npa{553}{675c}{1993}; e-Print
Archive: hep-ph/9210220
}
\REF\cat{
H.B. Nielsen, in the proceedings of the {\sl Workshop on Skyrmions and
Anomalies},  Cracow, Poland, Feb 20-24, 1987.  Edited by M. Jezabek and M.
Praszatowicz. (World Scientific, 1987)
H.B. Nielsen and A. Wirzba, in the proceedings of the {\sl Workshop on
Theoretical Physics: Elementary Structure of Material}, Les Houches, France,
Mar 24 - Apr 2, 1987. Edited by J.-M. Richard, E. Aslanides, N. Boccara.
(Springer-Verlag, 1988).
H.B. Nielsen \etal, \plb{281}{345}{1992}.
A. Wirzba, Ref. \bag}
\REF\gp{D.J. Gross \etal, Ref. \intro.}
\REF\extsource{
J.E. Mandula, \prd{14}{3497}{1976}.
E. Witten, \prl{38}{121}{1977}.
M. Magg, \plb{77}{199}{1978}.
P. Sikivie and N. Weiss, \prd{18}{3809}{1978}.
P. Pirila and P. Presnajder, \npb{142}{229}{1978}.
L. Jacobs and J. Wudka, \prd{25}{1114}{1982}.
H. Arodz, \app{14}{825}{1983}.
D. Sivers, \prd{34}{1141}{1986}.
E. Farhi \etal, \prd{47}{5551}{1993}.
E. Farhi \etal, \prd{50}{4162}{1994}.
}
\REF\jjr{
R. Jackiw \etal, \prd{20}{474}{1979}.
}
\REF\sphsym{E. Witten, \prl{38}{121}{1977}.}
\REF\rajeev{K.S. Gupta etal, \prd{48}{3354}{1993}.}
\REF\gluondet{Gross \etal, Ref. \intro.
N. Weiss, \prd{24}{475}{1981}; \prd{25}{2667}{1982}.}
\REF\wong{
S.K. Wong, \nci{65A}{689}{1970}.
S. Sternberg, \pnas{74}{5253}{1977}.
R. Giachetti \etal, \jmp{22}{1703}{1981}.
Ch. Duval and P. Horvathy, \ap{142}{10}{1982}.
See also
{\sl Gauge symmetries and fiber bundles: applications to particle
dynamics}; Lecture Notes in Physics, 188. Edited by A.P. Balachandran \etal
(Springer-Verlag, 1983).}
\REF\browei{ L.S. Brown and W.I. Weisberger, \npb{157}{285}{1979}; erratum-{\it
ibid.}
{\bf B172},{544},{(1980)}. }
\REF\llsm{L.D. Landau and E.M. Lifshitz, {\sl Statistical Physics} (3rd ed.),
(Pergamon Press, Oxford, New York, 1980)}
\REF\jr{R. Jackiw and P. Rossi, \prd{21}{426}{1980}.}
\REF\rhicref{
E.V. Shuryak, lecture presented at the {\sl
4th Workshop on Experiments and Detectors for a Relativistic Heavy Ion
Collider (RHIC)}, Upton, N.Y., Jul 1990. Edited by M. Fatyga and B. Moskowitz.
(Brookhaven National Lab, 1990, BNL-52262,C90/07/02.2).
D. Lissauer, in {\sl High energy nuclear collisions and quark
gluon plasma}, Kyoto 1991.
H. Satz, \npa{544}{371}{1992}.
J.W. Harris, in {\sl NATO Advanced Study Workshop on Hot Hadronic
Matter: Theory and Experiment}, Divonne-les-Bains, Switzerland,
27 Jun - 1 Jul 1994.
J. Schukraft, \npa{583}{673C}{1995}.
T.D. Lee, \npa{590}{11}{1995}.
}
\REF\zeroNatrhic{J.D. Bjorken, in {\sl Current Induced Reactions}, Lecture
Notes in Physics No. 56, edited by J.G. K\"orner  \etal (Springer,
Berlin), p. 93. See also L. McLerran, Ref. \intro\ and references
therein.}
\REF\collcoords{
G.S. Adkins \etal, \npb{228}{552}{1983}.
E. Guadagnini, \npb{236}{35}{1984}.
S. Jain and S.R. Wadia, \npb{258}{713}{1985}.
}
%

\advance \hoffset by 0.34 truein
\advance \voffset by -0.26 truein
\noblackbox

\def\thecaption#1{\global\advance\fignum by 1
\centerline{\hfil\vbox to 1 in{\hsize 5 in \vfil
{$\mskip -52.24mu$\textindent{\smallitii Fig:{\rm\ } \the\fignum{\rm\ }}
        \global \advance \baselineskip by -10 pt
        \smalli\noindent #1 }}\hfil}\global \advance \baselineskip by 10 pt
\bigskip}

\def\tf{Thomas-Fermi}
\def\ym{Yang-Mills}
\def\rh{\hat \rr }
\def\bo{\bcal_0}
\def\ai{\acal_1}
\def\ao{\acal_0}
\def\kt{ k \tcal }
\def\pkt{ \pi \kt }
\def\tf{Thomas-Fermi}
\def\tw{\lambda }
\def\dv{{\delta \vcal }}
\def\unitsPTV{($P$ in \mev$^4$, $V$ in fm$^3$, $T$ in \mev; the logarithms are
base 10).}
\def\unitsPVN{($P$ in \mev$^4$, $V$ in fm$^3$; the logarithms are base 10).}
\def\unitsPVmu{($P$ in \mev$^4$, $V$ in fm$^3$, $\mu$ in \mev; the logarithms
are base 10).}
\def\eden{\hbox{\sl e}}
\def\sden{\hbox{\sl s}}
\def\qden{\hbox{\sl q}}
\def\nden{\hbox{\sl n}}

\chapter{Introduction}

The equation of state of a quarks gluon plasma at high temperatures and/or
densities is one of the most important unknowns in our current
understanding of strong
interaction physics~\refmark{\intro}.
The applications of such an equation of state are
varied, ranging from cosmological compact objects to the physics of
heavy ion collisions. Unfortunately, due to the high degree of
non-linearity present in QCD, the determination of this equation of state
has proved to be a difficult task. For example, even within perturbation
theory, infrared singularities require the calculation of an infinite
number of graphs for the partition function beyond fifth
order~\refmark{\IRbehaviour}. General expressions for Green's functions
are available for the case where the internal momentum is large ($ \sim
\tcal $) while the external momenta are soft ($ \sim g \tcal $),
the so-called hard-thermal  loop region~\refmark{\htl}; using
standard manipulations~\refmark{\kapusta}, one can then determine the
partition function corresponding to all hard modes in the theory.
The soft-mode contributions to the partition function have been
studied using various approximations~\refmark{\soft} and numerical
calculations have also been developed (though not to the extent as
in the $ T = 0 $ case)~\refmark{\mc}.

In this paper we propose a new  approximation within which the physics of
a quark-gluon plasma can be studied. The formalism is based on the
\tf\ model of the atom~\refmark{\atomictf} and will be called \tf\ QCD
(TFQCD). We consider a plasma of quarks and gluons confined to a volume
$ \vcal $ which we imagine subdivided into a
number of subvolumes, each of which is large enough for the partons they
contain to be
considered a statistical ensemble. These subvolumes interact via a
background gauge field whose sources
are the thermally-averaged non-Abelian charge densities of the
subvolumes. The subvolumes are assumed to be small enough for the
background field to vary very little inside them, and because of this the
background field sources are essentially point-like. The requirement of
stability, together with the \ym\ equations for the
background field, furnishes a closed set
of equations which can be solved; from the solution the equation of
state for the system can be derived.
This program requires the evaluation of the thermally-averaged
non-Abelian charge densities which we obtain using perturbation theory;
in this paper we will use the lowest-order approximation, but a
systematic improvement is straightforward.

The atomic \tf\ approximation is useful when calculating bulk properties
of an atom with a large number of electrons, such as the total
ionization energy~\refmark{\bj}; it is also useful as a starting point for a
Hartree-Fock approximation. We expect the TFQCD model of a
quark-gluon plasma to be reasonably accurate for bulk properties of the
system, such as the equation of state. There are some differences
between the atomic \tf\ and the TFQCD formalisms; in particular note
that, in contrast to the atomic case, the quark-gluon plasma is not
stable: if left alone it will
fly apart and undergo a phase transition into a gas of hadrons. In order
to study a gas of quarks and gluons we are forced to imagine the system
to be enclosed in a container at sufficiently large temperature and/or
density.

The presence of an external confining agency is reminiscent of the bag
model~\refmark{\bag, \cat}. Through most of the paper we will consider, in
contrast
to the usual bag models, a situation where the partons are not confined,
and for which the external pressure is assumed to be generated by a some
physical apparatus. Despite this difference the bag boundary conditions
are also relevant for the present model: the system is assumed to be
confined to a spherical volume out of which neither fermion number nor
color can escape, this requires we impose both the original~\refmark{\bag}
and chiral~\refmark{\cat} bag boundary conditions.

We will also briefly study a system corresponding
to a hadron at zero temperature, and will show that the bag constant and
strong coupling constant obtained in the present approach are consistent
with those obtained using the bag model.

The volume of the system $ \vcal $ will be kept finite in all computations; the
results will then include finite-volume effects (such as terms in the
extensive thermodynamic quantities proportional $ \vcal^{2/3} $). In the
infinite volume limit these surface effects can be neglected and the
equation of state reduces to that of an ideal gas of gluons and quarks.

In the following section we will describe the construction of the TFQCD
model and present some simple applications. We will concentrate on the
case of an $ \su2 $ gauge theory with a single species of massless
fermions,
and then describe the modification required for the important case of
an $ \su3 $ gauge theory with three (massive) fermion flavors. The remainder
of the paper is organized as follows.
In section 3 we derive the equation of state for this model in the cases
of zero baryon number and zero temperature. The discussion of the
extension to $ \su3 $ and to more flavors is presented in
section 4. Some parting comments are presented in section 5; and a
mathematical detail is given in the appendix.

\chapter{Description of the model}

The model we propose is, as mentioned above, an extension of the \tf\
model to the case of QCD. We consider a gas of partons inside a volume $
\vcal $; we then imagine partitioning $ \vcal $
into small subvolumes $ \dv $ which are big enough so that the partons
(quarks and gluons) contained in them
form a statistical ensemble determined by a
temperature $ \tcal $ and, for the fermions, a chemical potential $ \mu
$. Each subvolume is required to be in equilibrium with its environment
which implies that the temperature and chemical potential are the
same throughout the system (this is intuitively obvious, we present a
proof in Sect. 2.3). The system is also assumed to be static so
that no currents are present.

We assume that the subvolumes have a non-zero average color charge, which
implies that the zero-component of the gauge field goes to a constant,
$ \bar A^0 $ at its boundary~\refmark{\gp}. We will refer to $ \bar
A^0 $ as the background gauge field. The background field is assumed to
vary slowly and smoothly between the $ \dv $, and is determined
self-consistently by requiring it to satisfy the \ym\
equations corresponding to the average charges of the subvolumes
(which themselves depend on the background fields). This approach
presupposes that the magnitude of the charge in any given $ \dv $
is small, and that the background field is approximately constant
within each subvolume; both these assumptions will be verified {\it
a-posteriori}.

Finally, we also assume that our system is spherically
symmetric; this requirement
considerably simplifies the calculations yet preserves the
essential non-Abelian character of the problem. The equations obtained for
the backgound fields are then similar to the ones derived when considering
the coupling of classical, spherically symmetric Yang-Mills fields to
external sources~\refmark{\extsource,\jjr}.

In the rest of this section we will treat the various ingredients of
the model separately. We first review the \ym\
equations within the spherically symmetric Ansatz.
We then obtain the expression for the partonic sources for the background
fields
and the various thermodynamic observables.
Next we derive the stability conditions for the
system.
Finally we combine these results in order to obtain the equations
for the background fields
which determine quantitatively the \tf-QCD (TFQCD) model.

The conventions which we use are the following.  The model is based on
an $ \su N $ \ym\ theory with one species of massless fermion;
the (anti-hermitian) group generators are denoted by $ T^a $ and the
gauge coupling constant by $g$. The covariant derivative is $ D_\mu =
\partial_\mu + A_\mu $ where $ A_\mu = g A_\mu^a T^a $. The
full Lagrangian is $$ \lcal = i \bar \psi \, \not \!\! D
\psi - \quarter \left( F_{\mu \nu } ^a \right)^2 \eqn\eq $$ where $ \psi
$ denotes the quark field, $A$ the gauge field, and
$$ F_{\mu \nu } ^a = \partial_\mu A_\nu^a -
\partial_\nu A_\mu ^a + g \epsilon_{ a b c } A_\mu^b A_\nu^c . \eqn\eq
$$ The sources are $$ j_\mu^a = i \bar \psi T^a
\gamma_\mu \psi . \eqn\eq $$ Latin indices from the beginning of the
alphabet ($ a , b , c $, etc) correspond to color indices; Latin indices form
the middle of the alphabet ($ i , j , k $, etc) denote space indices.

\section{Spherically symmetric gauge potentials and equations}

As mentioned in the previous section we will assume that the long-range
forces in our system are described by a non-Abelian background gauge
field generated by the average charge of each subvolume. We also
assume the system to have spherical symmetry. Thus we need the most
general expression for a spherically symmetric non-Abelian gauge field,
which is well known~\refmark{\sphsym}, and is reviewed for completeness below.

The most general spherically symmetric Ansatz for the gauge potentials
of an Abelian theory is simply $
A^0 = \phi (r,t)  $, $ \AA = a ( r,t ) \; \rh $, where \AA\ denotes
the vector potential and $r = | \rr |$. It is clear, however, that we
can choose a gauge where $ a ( r,t )  = 0 $, so we can take
$ \AA = 0 $.

 For the $ \su2 $ non-Abelian case the structure is much
richer~\foot{The situation is similar for larger groups, see section 4.};
the most general spherically symmetric Ansatz is~\refmark{\sphsym}
(the over-bar denotes the background fields)
$$ \eqalign{
\bar A_a^0 &= \ao \rh_a \cr
\bar A_a^i &= \epsilon_{ i a j } \rh_j \left( { g^{-1} + \varphi_2
\over r } \right) +  \left( \delta_{ i a } - \rh_i \rh_a \right) {
\varphi_1 \over r } + \rh_i \rh_a \ai \cr } \eqn\sphersymanz $$
which exhibits spin-isospin mixing~\foot{In this respect the
present approach differs from other investigations into spherically
symmetric hadron physics; see Ref. \rajeev.}.
The fields $ \varphi_{ 1 , 2 } $ and $ \acal_{ 0 , 1 } $ depend on
$r$ and $t$.

Within this Ansatz the  $ \su2 $ \ym\ Lagrangian becomes
$$ \quarter F_{ \mu \nu }^a {}^2  =
\quarter f_{ \mu \nu } ^2 - \inv{ r^2 } \left|
\dcal \Phi \right|^2 + { g^2 \over 2 r^4 } \left( \left| \Phi \right|^2 -
\inv{ g^2 } \right)^2 \eqn\sphersymlagr $$ where the indices $ \mu, \nu , $
etc. equal $1$
(corresponding to $r$) or $0$ (corresponding to $t$); the metric is
diag$( 1 , - 1) $. We also defined $ f_{ \mu \nu } = \partial_\mu \acal_\nu
- \partial_\nu \acal_\mu $, $\Phi = \varphi_1 + i \varphi_2 $ and $
\dcal_\mu = \partial_\mu + i g \acal_\mu $. The above
expression is invariant under the gauge transformation $$ \acal_\mu \rightarrow
\acal_\mu - \partial_\mu \Lambda , \qquad \Phi \rightarrow e^{ i g \Lambda }
\Phi ,\eqn\sphsymgaugetransf $$ which is a remnant of the original
non-Abelian invariance.

We now consider the
coupling of the above fields to a spherically symmetric charge
density $ \rho^a $, where spherical symmetry requires $ \rho^a = \qden
\hat r^a $. The coupling is then described by adding a term $$ \lcal_{
\rm interaction} = g \bar A_0^a \rho^a = g \, \qden \, \ao
\eqn\sphersymint $$ to the Lagrangian.

\section{The partition function for gluons and fermions}

We now imagine that the
volume of the system, denoted by $ \vcal $, is
subdivided into a large number of subvolumes $
\dv $. The gauge fields inside each subvolume are separated into a
background piece $ \bar A_\mu^a $ and a fluctuation $ a_\mu^a $: $
A_\mu^a = \bar A_\mu^a + a_\mu^a $.

In this subsection we evaluate the partition function for
the partons inside $ \dv $. This object, which we call $ Z_\dv $, will
depend on $ \bar A_\mu^a $, and we can use this dependence to obtain the
thermal average of the non-Abelian currents, $$ \bar j_\mu^a = \inv g
\deriva{ Z_\dv }{ \bar A^\mu{}^a }{} \eqn\eq $$

Since the system is supposed to be in a static configuration we require $
\bar j_i^a = 0 $ which implies we can take $ \bar A_i^a = 0 $ inside $ \dv $.
Since the background fields are assumed
slowly varying, we also take $ \bar A^a_0 $
constant inside $ \dv $; we then choose a gauge such
that $ \bar A^a_0 $ is diagonal inside $ \dv $.
Hence $Z_\dv$ will depend  on the temperature $ \tcal $, the
fermionic chemical potential $ \mu $, and
the $n$ components of $ \bar A_0 $ associated with the
diagonal generators ($n$ is the rank of the gauge group).

As a first approximation we will neglect the interaction
between the fermions and the $ a_\mu^a $, as well as  the non-linear
couplings among the $ a_\mu^a $; these interactions can be included
perturbatively.

Concerning the scale of $ \dv $ we will assume that it is set by the
fermion thermal wavelength, $ \tw $, that is, $$ \dv \sim
\tw^3 . \eqn\eq $$

We will for the moment
restrict ourselves to the case where the gauge group is $
\su2$ (the extension to $ \su3 $ will be described in section 4
below). In this case
the group generators are $ T^a = \sigma^a/ ( 2 i ) $ and, within $ \dv
$,  $ \bar A_0 = g \bar A_0^a T^a = g \ao \sigma_3/ ( 2 i ) $, so that
$$ A^b_0 = a^b_0 + \ao \delta_{b 3}, \qquad
   A_i^b = a^b_1 \quad (\hbox{inside } \dv) \eqn\gluonsindv $$

We first evaluate the fermionic contribution to the partition
function, and then calculate the contributions
from the $ a_\mu^a $.

\subsection{Fermionic contribution.}

When considering the fermionic partition function we
will assume only one massless fermions species (the modifications required
by several species and/or non-zero masses are straightforward). Thus
we look for an approximate expression for $ Z_\psi =  \det \left( i \not
\!\! D + \mu \gamma_0 \right)$ where the gluon fields take the form
\gluonsindv. Inside a subvolume $ \dv $ it is
assumed that the fermions behave as a statistical ensemble,
that the interaction with the $ a_\mu^a $ is small, and that the
background fields $ \bar A_\mu^a $ are essentially constant.
Adopting these approximations we reduce the calculation to evaluating
$ \det \left[ i \not \! \partial + \left(-  i \bar A_0 + \mu \right)
\gamma_0
\right] $ with $ \bar A_0 = g \ao \sigma_3/(2 i ) $,
$ \acal_ 0 =$constant.

The partition function for an ideal gas of massless fermions at
temperature $ \tcal $ in a volume $ \dv $ and with a chemical
potential equal to $ \mu $ is given by~\refmark{\kapusta} $$
\ln Z_0 = { \beta \, \dv \over 12 \pi^2 }
\left[ \mu^4 + 2 ( \pkt )^2 \mu^2 + {
7 \over 15 } ( \pkt )^4 \right] \eqn\freeZ$$ where the zero
subscript indicates that no gauge fields are included.
The constant background gauge fields
are then included by replacing $ \mu \rightarrow
 \mu \pm g \ao/2 $ in $ Z_0 $ with the sign depending on the
isospin of fermion, and adding the contributions from each isospin
component. Thus, within the above approximations, we obtain
$$ \ln Z_\psi  = { \beta \dv \over
6 \pi^2 } \left[ { g^4 \ao^4 \over 16 }  +
{ g^2 \ao^2 \over 8 \tw^2 } + \mu^4 + 2 \left(
\pkt \mu \right)^2 + { 7 \over 15 } \left( \pkt \right)^4 \right]
\eqn\defofzpsi
$$ where we defined the thermal wavelength $$ \tw = \inv{ 2 \sqrt{ 3
\mu^2 + \left( \pkt \right)^2 } } \eqn\defoftw $$

This approximation to $Z_\psi$ generates the following expressions for the
fermionic contribution to the (local) thermodynamic quantities
$$ \eqalign{ \pcal_\psi =& { \ln Z_\psi \over \beta \dv } =
\inv{ 6 \pi^2 } \left[ { g^4 \ao^4 \over 16 }
+ { g^2 \ao^2 \over 8 \tw^2 } + \mu^4 + 2 \left(
\pkt \mu \right)^2 + { 7 \over 15 } \left( \pkt \right)^4 \right] , \cr
\sden_\psi =& { 2 k \over 3 \pi } ( \pkt ) \left[
{ g^2 \ao^2 \over 4 } + \mu^2 + {7 \over 15 } ( \pkt )^2 \right] , \cr
\eden_\psi =& \inv{ 2 \pi^2 } \left[
{ g^2 \ao^2 \over 24 \tw^2  } -
{ g^4 \ao^4 \over 48 } + \mu^4 + 2 ( \pkt \mu )^2  + { 7 \over 15 } (
\pkt )^4 \right] \cr
\nden =& \inv{ \beta \dv } \deriva {\ln Z_\psi } \mu {} =
{ 2 \mu \over 3 \pi^2 } \left[ { 3 \over 4 } g^2 \ao^2 +
\mu^2 + \left( \pkt \right)^2
\right] , \cr
\qden_\psi =& \inv{ g \beta \dv }  \deriva {\ln Z_\psi } \ao {} =
{ g \ao \over 24 \pi^2 } \left[ g^2 \ao^2  + \inv{ \tw^2 }
 \right] \cr } \eqn\fermionstuff $$
where $ \pcal $ denotes the pressure and \sden, \eden, \nden, \qden, etc denote
the entropy, energy, particle and charge {\it per unit volume}.
Note that $ \eden_\psi = 3 \pcal_\psi - 2 \ao \qden $ includes the energy of
the interaction with the gauge fields.

\subsection{Gluonic contribution}

The gluonic contribution to the partition function
is obtained in a manner similar to the one
followed for the fermions. As before we will ignore the
self-interactions of the fields $ a_\mu^a $, in this case the partition
function, including the Fadeev-Popov determinant, reduces to $ Z\lowti{gluons}
=
\det{- \bar D\lowti{adj}^2} $ where
$\bar D\lowti{adj}$ denote the covariant derivative for the background
fields in the adjoint representation~\refmark{\gluondet}.

In calculating this determinant we take into account the partons are
supposed to be in a box of side $ \sim \tw $. Moreover, the background
field is supposed to include the effects from the zero (Fourier) modes
in the field. It follows that we need to include only modes with energy
above $$ p_o = { 2 \pi \over \tw } . \eqn\eq$$ Using the
gauge $ \bar A_\mu^a = \delta_{\mu , 0} \delta^{a,3} \ao $ gives
$$ \ln Z\lowti{gluons} = - \vcal \int_{ p > p_o} { d^3 p \over ( 2 \pi )^3 }
\left[
\ln \left( 1 - e^{ - \beta ( p - g \ao ) } \right) +
\ln \left( 1 - e^{ - \beta ( p + g \ao ) } \right) +
\ln \left( 1 - e^{ - \beta p } \right) \right] \eqn\gluonz $$
which corresponds to a gas of massless bosons with chemical potential $
\pm g \ao $ and $0$.

We will argue below (section 2.4.1) that the background
field $ \ao $ is monotonic in $r$ and that $ \tw \, g \ao \le 3 \pi /2
$. Using also the fact that $ \beta p_o \ge 4 \pi ^2 $ we find that to a
good approximation $$
\pcal \lowti{gluons} = { \left( \pkt \right)^4 \over 15 \pi^2 } + {4 \left( k T
\right)^4 \over \pi^2 } \left( \beta^2 p_o^2 + 2 \beta p_o + 2 \right)
e^{- \beta p_0} \, \left[ \sinh \left( \beta g \ao /2 \right) \right]^2 \eqn\eq
$$
This shows that the
deviations from the free-gluon values are exponentially suppressed
(recall that $ \beta p_o \ge 4 \pi^2 $)  and can be neglected. In this case
$$ \matrix{
\pcal \lowti{gluons} \simeq { \left( \pkt \right)^4 \over 15 \pi^2 } \hfill &
\eden \lowti{gluons} \simeq { \left( \pkt \right)^4 \over 5 \pi^2 } \hfill \cr
\sden \lowti{gluons} \simeq { 4 k \left( \pkt \right)^3 \over 15 \pi} \hfill &
\qden \lowti{gluons} \simeq 0 \hfill \cr }
\eqn\eq $$
where $ \pcal $ denotes the pressure, and \eden, \sden\ and \qden the
energy, entropy and charge per unit volume respectively. The errors
incurred are below a few percent for the thermodynamic quantities and
below 0.0075\% for the charge.

\section{Stability conditions}

The stability criterion can be obtained from the Wong
equations~\refmark{\wong}, but a more elegant
argument can be gleaned from a paper by Brown and
Weisberger~\refmark{\browei}.
Consider the background field contribution to the energy momentum tensor $
\theta\lowti{back} $, which satisfies
$$ \partial_\mu \theta^{ \mu \nu }\lowti{back} = g
\rho_a \bar F_a^{ \nu 0 } . \eqn\eq $$
where $ \bar F_a^{ \nu 0 } $ denotes the field strength for the
background fields, and $ \rho_a$ the thermally-averaged
non-Abelian charge density.

Since the total energy momentum tensor is conserved,
it follows that the averaged partonic contribution $ \theta\lowti{part} $
satisfies $$
\partial_\mu \theta^{ \mu \nu } \lowti{part} =
- g \rho_a \bar F_a^{ \nu 0 } . \eqn\eq $$ For static
situations the above equation implies $$
\partial_i \theta^{ i j } \lowti{part} = - g \rho_a \bar F_a^{ j 0 } .\eqn\eq
$$
If in addition
we impose spherical symmetry (see section 2.1) $ \rho_a \propto \hat r^a $,
$ \bar A_0^a \propto \hat r^a $ which
implies $ \rho_a \bar F_a^{ j 0 } = - \rho_a \partial_j \bar A_0^a $.
For a homogeneous gas of partons the space components of the
energy momentum  tensor are $
\theta\lowti{part}^{ i j } = \pcal\lowti{part} \delta_{ i j } $.
Collecting these results we get $ \partial_j \pcal\lowti{part} =
g \rho_a \partial_j \bar A_0^a $, or, equivalently $$
d \pcal\lowti{part} = g \rho_a \, d \bar A_0^a \eqn\stabcond $$
which is the desired constraint.

For the Abelian case \stabcond\ reduces to the usual Thomas-Fermi
equilibrium condition: the pressure on $ \dv $ is balanced by the
electrostatic force.

This stability condition requires the chemical potential and temperature
to be $r$ independent.
Indeed, $ \ln Z_\dv $, the parton partition function for a small volume
$ \dv $, is a function of $\tcal $, $\ao $ and
$\mu $; using \fermionstuff\ we obtain
$$ \eqalign{ d \pcal_{\rm part} =&
\inv{ \beta \vcal } \deriva { \ln Z_\dv } \ao {} \, d \ao +
\inv{ \beta \vcal } \deriva { \ln Z_\dv } \mu {} \, d \mu +
\inv{ \beta \vcal } \deriva { \ln Z_\dv } \tcal {} \, d \tcal \cr
=& g \qden \, d\ao + \nden_{\rm part} \, d\mu + ( \eden_{\rm part} - \mu
\nden_{\rm part} ) { d \tcal \over \tcal } \cr } \eqn\eq $$ where $ \nden_{\rm
part} $
and $ \eden_{\rm part} $ are, respectively, the particle and energy
densities of the
partons, and $ \rho^a = \qden \hat r^a $. Substituting this expression
for $ \rho_a $, using \sphersymanz, and comparing to \stabcond\
we obtain  $ d \mu =  d \tcal =0 $.

\section{The TFQCD equations}

The equations of motion are derived from the spherically symmetric
Lagrangian for the background fields
\sphersymlagr\ when the potentials interact with a
source \qden\ according \sphersymint. The source, given in \fermionstuff,
is itself a  function of the potentials. The resulting equations are
$$ \eqalign{ \dcal^2 \Phi + { g^2 \over r^2 } \left( \left| \Phi
\right|^2 - \inv{ g^2 } \right) \Phi & = 0 \cr
\partial^\mu \left( r^2 f_{ \mu \nu } \right) + 2 g \im \left( \Phi^*
\dcal_\nu \Phi \right) & = -g r^2 \qden \delta_{ \nu , 0 } . \cr } \eqn\eq
$$ The gauge invariance of these equations allows us to chose the $
\acal_1 = 0 $ gauge. The second of the above equations gives, when $ \nu
= 1 $ and for static configurations, $ \im \Phi^* \Phi' = 0 $, so that we
can choose $ \Phi $ to be purely imaginary.

We will use the notation~\refmark{\extsource,\jjr}
$$ \ao= { f (r) \over r g } , \qquad \Phi =
\inv{ i g }  a (r) . \eqn\afdef $$
Then, using \fermionstuff, the above equations become
$$ \eqalign{ f '' -
2 \left( { a \over x } \right)^2 f =& { \alpha \over 6 \pi } f
\left( { f^2 \over x^2 } + 1 \right) \cr a '' + { 1 + f^2 - a^2
\over x^2 } a =& 0 . \cr} \eqn\eom $$ where $ \alpha = g^2 / ( 4 \pi ) $
and $ x = r / \tw $ with $ \tw $ defined in \defoftw.

These equations determine the background self-consistently. Their
solution requires the specifications of the boundary conditions to which
we now turn.

\subsection{Boundary conditions}
The conditions near the origin are determined by considering the
behavior of Wilson loops as $ r \rightarrow 0 $.
We find that singularities arise
unless $f$ and $ a^2 -1 $ vanish at $ r =0 $. Using the
$ a \leftrightarrow - a $ symmetry (which is a remnant of the
gauge symmetry) we can then require $ a\rightarrow 1 $ as $ r
\rightarrow 0 $. The precise manner in which $f$ and $ a -1 $ vanish as
$r$ approaches zero is determined by requiring that the energy should
have no divergences at this point. We then obtain $$ f , \ a-1 = O ( r^2 )
\quad \hbox{for\ }  r \rightarrow 0 . \eqn\bco $$

It is easy to see that the equations of motion \eom\ require
$f$ to be concave or convex; since we can always
exchange $ f \rightarrow -f $ we can assume that $f$ is
concave. In particular this implies that $f$ will not vanish for $ r > 0
$. From \eom\ it is also easy to show that $ ( f/ r )' > 0 $.
In contrast $a$ can (and will) have extrema as well as zeroes.

As mentioned in the introduction we assume that the system is enclosed
in a container which must be spherical due to  requirement of spherical
symmetry; we denote by $R$ its radius. If the system is to
be confined to the region $ r < R $, there should be no leakage of fermion
number or color into the region $ r > R $.

The first of these two conditions (fermion number confinement)
requires the fermions to
have zero radial component of the momentum at $ r = R $. This implies that
in the vicinity of $ r = R $ the fermion gas becomes two dimensional.
The corresponding (surface) charge density $ \sigma^a $ takes the form
$$ \sigma^a = \vartheta \, \hat r^a \eqn\eq $$ as mandated by spherical
symmetry. Note however that $ \vartheta $ does not
have a simple analytical form, $$ \eqalign{ \vartheta &=  \inv{ 8 \pi
\beta^2 }  \left[ \qcal ( \beta \mu + \beta g \ao / 2 ) -  \qcal ( \beta
\mu - \beta  g \ao / 2 ) \right] \cr \qcal ( u ) &= 2 \int_0^\infty ds
\, \ln \left[ { e^s + e^u \over e^s + e^{ - u } } \right]
\simeq
\left( u^2 + { \pi^2 \over 3} \right) \tanh \left( { 12 \ln 2 \over
\pi^2 } \, u \right). \cr }
\eqn\eq $$ where the analytic approximation to $ \qcal $ is accurate to
about $ 0.62 \% $; the derivative is accurate to $ 0.92\% $.

In the examples which we consider in detail we
will be interested in the limit where $ \ao $ is large and where $
\beta \rightarrow 0 $ or $ \mu \rightarrow 0 $. In these cases we have
$$ \vartheta \simeq { g^2 \over 16 \pi } \ao^2 . \eqn\eq $$

We will
require the volume charge density in the bulk to smoothly join the
surface charge density at the surface layer, that is, $ \rho^a \tw
= \sigma^a $ at $ r= R $. Thus we impose, $ \qden \tw = \vartheta $
at $ r = R $ which, keeping in mind that the solutions produce large
values of $ \ao $ at $R$, is equivalent to $ \ao^2/ ( 4 \pi ) =
\tw \ao^3 / ( 3 \pi^2 ) $, or equivalently $$ f ( R ) = { 3 \pi
\over 2 } { R \over \tw } \eqn\bci $$ It is of course possible to modify
this condition by requiring only that, at $ r = R $, $ \qden \tw = s
\vartheta $ for some number $ s = O ( 1 ) $, which is equivalent to
replacing $ \tw \rightarrow \tw / s $; our results are
insensitive to such a replacement.

To determine the consequences of the second of the above two conditions
(color confinement)
we need the components of the chromo-electric and chromo-magnetic
fields parallel and perpendicular to \rr,
$$ \eqalign{
\hat \rr \cdot \EE^a &= - \inv{ g^2 \tw } \left( {f \over x } \right)'  \hat
r^a \qquad
 \hat \rr \cdot \BB^a = - \inv{ g^2 \tw } { a^2 - 1 \over x^2 } \hat r^a \cr
\left( \hat \rr \times \BB^a \right)^i &= - \inv{ g^2 \tw }
{ a' \over x } \epsilon_{ i j a } \hat r^j \qquad
\left( \hat \rr \times \EE^a \right)^i = - \inv{ g^2 \tw }
{ f a \over x^2 } \epsilon_{ i j a } \hat r^j \cr } \eqn\eq $$
The first of these relations, together with the previously derived
result $ ( f / r ) ' \not= 0 $, implies that color
will leak from the system unless an appropriate modification is included.
The situation is identical to the one present in the bag model~\refmark{\bag},
and the
solution which we adopt is the same~\refmark{\cat}.
We will couple our system at the $ r = R $ boundary
to a CP-odd field $ \eta' $ via a term proportional
to the Chern-Simons term; this coupling insures that color is confined
to the region $ r \le R $~\refmark{\cat}.
Denoting by $ F_{ \eta' } $ the decay constant on the $ \eta' $, the
coupling to this field at $ r = R $ are determined by the relations
$$ \rh \cdot \EE^a = { \alpha \over \pi F_{ \eta' } } \rh \cdot \BB^a \;
\eta' ; \qquad \rh \times \BB^a = - { \alpha \over \pi F_{ \eta' } } \rh
\times \EE^a \; \eta'  \eqn\eq $$ from which we derive
$ ( \hat \rr \cdot \EE^a )  \left( \hat \rr \times \EE^a \right) +
( \hat \rr \cdot \BB^a ) \left( \hat \rr \times \BB^a \right) = 0 $; in
terms of the $a$ and $f$ fields this becomes $$ f a ( x f' - f ) + x a'
( a^2 - 1 ) =0 \quad \hbox{at} \ r=R \eqn\bcii $$ which is the desired
condition.~\foot{Concerning \bcii\ we know, from the numerical
integration of \eom,
that $ f ( x f' - f ) $ does not vanish, we also find that it is
numerically large for the situations we consider in detail.
It follows that \bcii\
can be approximately replaced by the simpler condition $ a( R )  = 0 $.}

\subsection{Character of the solutions}

The TFQCD potentials $f$ and $a$ are then obtained by solving the
equations \eom\ subject to the boundary conditions \bco, \bci\ and \bcii.
These solutions, as well as all thermodynamic variables, will depend on the
parameter $ \alpha = g^2 /  ( 4 \pi ) $. In order to specify $ \alpha$
we first fix the thermodynamic variables of the system, such as the
energy and volume; the TFQCD expresses these thermodynamic variables as
functions of $ \alpha $, which is chosen so that the chosen values are
met.

When considering the \eom\ we find that, for given
values of $X$ and $ \alpha $, there are several solutions satisfying the
boundary conditions~\foot{This is reminiscent of the situations found in the
case of classical solutions to the \ym\ equations with external
sources~\refmark{\jjr}).}. Of these solutions there is a set
(we, in fact, found two such solutions) which minimizes
$ \Omega $, the thermodynamic potential at constant pressure and chemical
potential~\refmark{\llsm}, $$ \Omega = - \int d^3 x \; \pcal , \eqn\eq$$ where
$
\pcal $ denotes the total pressure.
Numerical studies show that there is no cross-over as $ \alpha $
changes: each member of the set of solution which minimizes
$ \Omega $ is a smooth
function of $ \alpha $. Selecting the solution which minimizes $ \Omega $ we
then
determined $ \alpha $ by matching the specified energy and baryon
number.

The explicit expressions for $ \Omega $, the total energy $ \ecal $ and
the total number of particles (baryon number) $ \ncal $ are $$ \eqalign{
\Omega =&  \inv{ \alpha \tw } \int_0^X dx \left\{
\half \left( f' - { f \over x } \right)^2 +
\half \left( { 1 - a^2 \over x } \right)^2 -
\left( a' \right)^2 -
\left( { f a \over x } \right)^2 - \right. \cr & \left.
- { \alpha \over 24 \pi } f^2 \left( 2 + { f^2 \over x^2 } \right) \right\}
- { 2 \over 9 \pi} \left[ \mu^4 + 2 \mu^2 \left( \pkt \right)^2
+ { 13 \over 15 } \left( \pkt \right)^4 \right] R^3 \cr } \eqn\pot $$
$$ \eqalign{
\ecal =& \inv{ \alpha \tw } \int_0^X dx \left\{
\half \left( f' - { f \over x } \right)^2 +
\half \left( { 1 - a^2 \over x } \right)^2 +
\left( a' \right)^2 +
\left( { f a \over x } \right)^2 + \right. \cr & \left.
+ { \alpha \over 24 \pi } f^2 \left( 2 - { f^2 \over x^2 } \right) \right\}
+ { 2 \over 3 \pi} \left[ \mu^4 + 2 \mu^2 \left( \pkt \right)^2
+ { 13 \over 15 } \left( \pkt \right)^4 \right] R^3 \cr } \eqn\energy $$
$$ \ncal = { 2 \mu \over \pi } \left\{
{ 4 \over 9 } \left[ \mu^2 + \left( \pkt \right )^2 \right] R^3
+ \tw \int_0^X dx \; f^2 \right\} , \eqn\partnum $$ where $$ X = { R
\over \tw } . \eqn\eq $$
For future reference we also provide the expression for the (total)
entropy of the system $$ \inv k \scal = {2 \over 3 } \tw \pkt
\left\{ { 4 \over 3 } X^3 \tw^2 \left[ \mu^2 + { 13 \over 15 }
\left( \pkt \right)^2 \right] +  \int_0^X f^2 \, dx \right\}
\eqn\theentropy $$

Solutions of the equations for $f$ and $a$ can be obtained using
standard numerical algorithms; due to the singular nature of the
equations at the origin the relaxation method is best suited.

\setbox2=\vbox to 270pt{\epsfxsize=7.5 truein\epsfbox[0 -230 612 562]{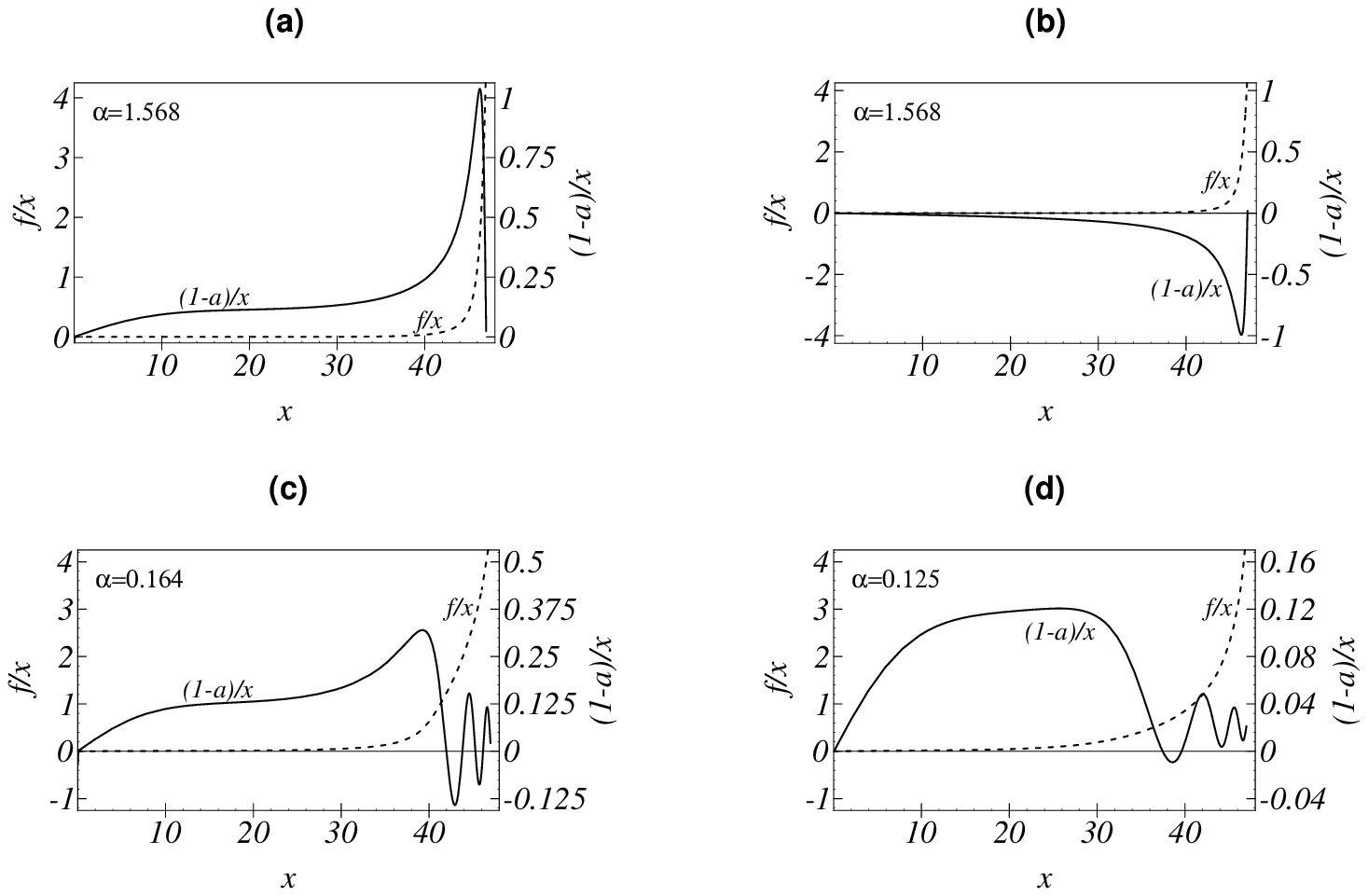}}
\centerline{ \box2 }

\thecaphyz{Examples of solutions $ f(x) $ (dashed lines) and $ a(x) $
(solid lines) corresponding to $R=10$fm, $ \ecal/\vcal
=4 \gev /fm^3 $ and $ \tcal =150 \mev$ ($X=47$).
Cases $(a) $ and $(B)$ corresponds to the solutions which minimize the
thermodynamic
potential. Cases $(c)$ and $(d)$ have larger $ \Omega $ and represent unstable
solutions; the values of $ \alpha $ corresponding to each solution are
indicated.}

We present several examples of the solutions in
\FIG\myfi{Examples of solutions $ f(x) $ (dashed lines) and $ a(x) $
(solid lines) corresponding to $R=10$fm, $ \ecal/\vcal
=4 \gev /fm^3 $ and $ \tcal =150 \mev$ ($X=47$).
Cases $(a) $ and $(B)$ corresponds to the solutions which minimize the
thermodynamic
potential. Cases $(c)$ and $(d)$ have larger $ \Omega $ and represent unstable
solutions; the values of $ \alpha $ corresponding to each solution are
indicated.}
Fig. \myfi\ where we took  $R=10$fm, $ \ecal/\vcal
=4 \gev /fm^3 $ and $ \tcal =150 \mev$ (which implies $X=47$).
All the solutions in Fig. \myfi\ satisfy the equations \eom\ and the
boundary conditions; the \
solutions which minimize the thermodynamic potential corresponds to
cases
(a) and (b)~\foot{The singular nature of the equations allows for the
multiplicity of solutions; we have found 8 solutions in total
(for the given values of $ \ecal $ and $ \vcal $ but having different values of
$ \alpha $),
though we cannot assert that this an exhaustive list. Using the
relaxation method, the solution that minimizes $ \Omega $ was lest sensitive
to the initial trial functions, solutions with larger $ \Omega $ become
increasingly more difficult to find as the range of initial
configurations which relax to such solutions of \eom\ becomes more and
more restricted. We have not attempted to perform an complete study
of the properties and number of solutions restricting ourselves to
finding the one solutions relevant for physical applications together with
some examples of unstable solutions.}. These two solutions correspond to
indistinguishable thermodynamics (within numerical errors); for the
calculations below this duality  presents no complications. We have not
attempted to study the stability of these solutions against non-radially
symmetric perturbations~\refmark{\jr}.

Given these
results we must now determine whether they are consistent with the
original assumptions, that is, whether $f$ varies slowly enough to be
considered constant in a region of width $ \sim \tw $. We also must
determine to what extent are color charges screened. The plots presented
correspond to both cases (a) and (b) in Fig. \myfi.

\setbox2=\vbox to 135pt{\epsfysize=11 truein\epsfbox[20 -300 632 492]{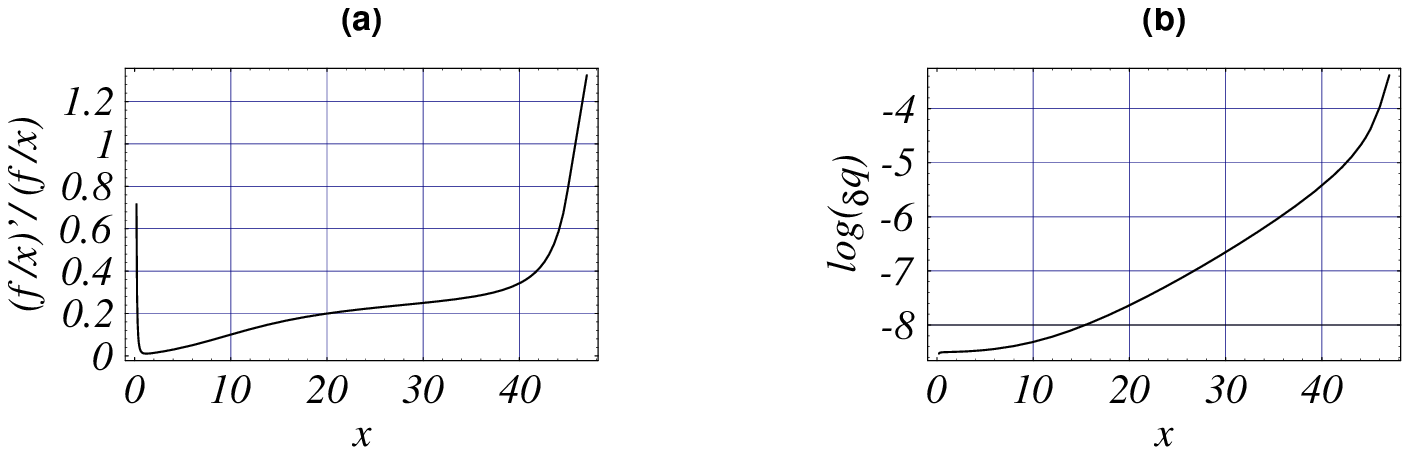}}
\centerline{ \box2 }
\thecaphyz{Validity of the \tf\ approximation. (a): only in the regions
near the boundary at $ x = 47 $ ($ x > 45.5 $) and the origin ($ x < 1
$) the approximation does breaks down. (b): charge is effectively screened
throughout the volume (the logarithm is base 10).}


The rate of change of $f$ is sufficiently slow provided the potential $ \ao $
changes little within a region of size $ \tw $, this is equivalent
to $$  {  ( f / x ) ' \over ( f / x )  } <  1 ; \eqn\tfqcdcond $$ a plot of
the \lhs of this equation if presented in
\FIG\myfii{Validity of the \tf\ approximation. Only in the regions
near the boundary at $ x = 47 $ ($ x > 45.5 $) and the origin ($ x < 1
$) the approximation does breaks down.}
Fig. \myfii(a).
We see that the condition \tfqcdcond\ is satisfied except in the
vicinity of the origin and the $ r = R $ boundary. The value of
\tfqcdcond\ near $ x = 0 $ presented in Fig. \myfii(a) is an underestimate
generated by numerical errors (the equations are singular at $ x = 0 $);
for $ x \rightarrow 0 $, $ ( f/x)' / ( f/x) \simeq 1/x $.

The magnitude of the charge in a subvolume $\dv \sim \tw^3 $
is obtained from \fermionstuff, and equals $$ \delta \qden = \tw^3 \,
\qden = \inv{24 \pi^2 } { f \over x } \left( { f^2 \over x^2 } + 1
\right), \eqn\defofdq $$ A plot of this quantity is presented in Fig.
\myfii(b); as can be seen the magnitude of the color charge inside each
subvolume is quite small except near the $ r = R $ boundary: the system does
screen
its charges quite effectively.

\section{Solutions for small $X$}

When $X$ is small then $f$ will be small also since it is monotonic
(this follows from the boundary condition \bci).
In this case the equation for $a$ decouples and so does the boundary
condition \bcii, $$ x^2 a'' + ( 1 - a^2 ) a = 0 ; \qquad a(0)=1, \
a'( X ) \left[ a(X)^2-1 \right] = 0 \eqn\eq $$ If we define
$$ a_2 = \half a''(0) , \eqn\eq $$
it is easy to see that the solution to the above equation is a
function of $ a_2 x^2 $. It is then enough to assume $ a_2 =\pm1 $; the
general solutions are obtained from these by rescaling $x$.
The solutions to the above differential equation (for $ a_2 =
\pm1 $ ) are presented in
\FIG\myfigpert{Solutions for small $X$, solid curve: $ a_2=-1 $, dashed
curve: $a_2= 1$.}
Fig \myfigpert. The solutions are monotonic, so the boundary condition
at $x=X$ is satisfied when $ a(X)=-1 $ which occurs only for
$ a_2 < 0 $, numerically $$ a_2 X^2 \simeq -4.1 \eqn\eq $$
which completely specifies the solution.

\setbox2=\vbox to 100pt{\epsfysize=4.7 truein\epsfbox[0 -200 612 592]{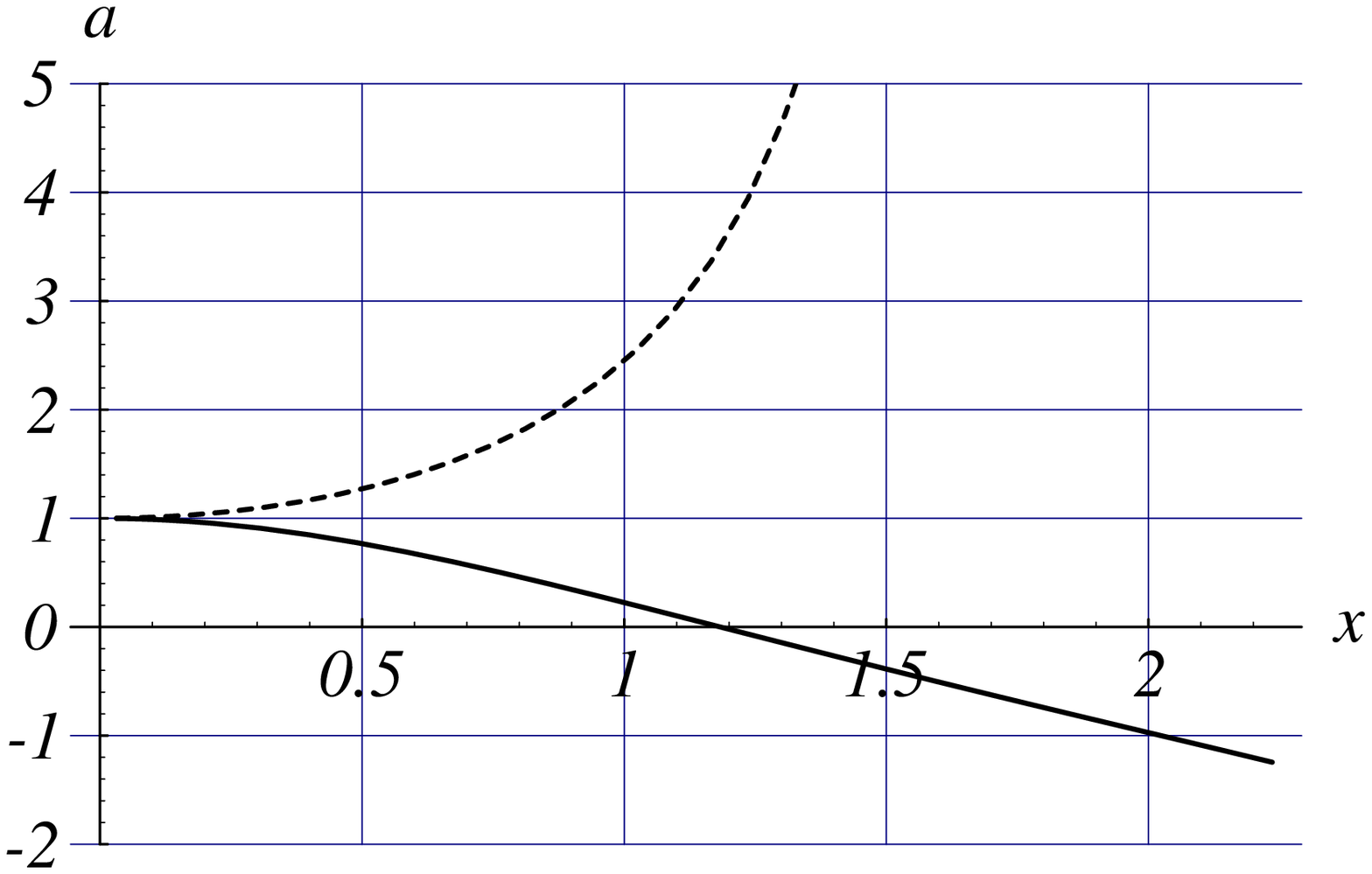}}

\centerline{\box2}
\thecaphyz{Solutions for small $X$, solid curve: $ a''(0)=-1 $, dashed
curve: $a''(0)= 1$.}

Again neglecting $f$ and evaluating numerically
the integrals gives
$$ \eqalign{
\ecal &\simeq { 5.42 \over \alpha R} + 4.28 ( \kt )^4 \vcal \cr
\pcal & \simeq {2.91 \over \alpha } \vcal^{-4/3} + 1.43 ( \kt )^4 \cr }
\eqn\eq $$
For example, at $ \tcal = 0 $, $ \ecal = 1 \gev $, $ R = 1$fm,
$ \alpha \simeq 1 $ and $ \pcal \simeq 85 \mev/ \fm^3$.
Fixing $\vcal$ and $\tcal $, the coupling strength
$ \alpha $ drops as $ 1 /\ecal $.

At zero temperature we have $ \ncal \simeq 2 \mu^3 \vcal/(3 \pi^2 ) $
and $$ \ecal \simeq { 5.42 \over \alpha R } + { \mu^4 \vcal \over 19.74}
, \qquad \pcal = { 2.91 \over \alpha } \vcal^{-4/3} { \mu^4 \over 59.04}
\eqn\eq $$ so that the equation of state becomes $$ \pcal \vcal^{4/3}
\simeq { 2.91 \over \alpha} + { \ncal^{4/3} \over 1.62 } \eqn\eoslowX $$
In this case $ \ncal = 2 $, $ \ecal = 1 \gev $ and $ R = 1 \fm $ imply $
\alpha \simeq 2.4 $ and $ \pcal \simeq 77 \mev/ \fm^3  $.

The numbers obtained for the case of small $X$ are then quite consistent
with those obtained using the bag model~\refmark{\bag}
(except perhaps for a large value for $ \alpha $). Note however that in the
present calculation the contributions from the non-ideal gas terms are
very important and the numerical agreement is not trivial. It is also
true that the present model is far from realistic (being based on an $
\su2 $ gauge theory with a single species of massless quarks). These
results are therefore quite encouraging but not conclusive as to the
physical relevance of this model.

Concerning the other thermodynamic quantities they relapse to their
free-particle values up to $ O ( f^2 ) $ corrections. Note that the adiabats
are, in general, defined by $ X = $const. which, for the case
$ \ncal = 0 $, imply $ \pcal^3 \vcal^4 = $const. just like a
relativistic ideal gas.

When $ \mu = 0 $ an
approximate solution for $f$ which satisfies the boundary conditions
is $$ f \simeq 4.57 { x^2 \over X } , \eqn\eq $$ in this case the
entropy becomes $ \scal \simeq 1.5 k X^3 $ and the heat capacity equals
$ C_V \simeq 3 \scal $; the largest contribution to these quantities
($\sim 94\%$) comes from the $ \int f^2 $ term.

\section{Solutions for large $X$}

In order to study the solutions to \eom\ for $x$ finite but
$ X \rightarrow \infty $ it proves convenient to define $ y = x/ X
$. We are then interested in the small $y$ behavior of the solutions and
a power series is appropriate,
$$
\eqalign{
f &=f_2 \left[ y^2 +\left( {2 a_2 \over 5} + u \right) y^4 + \left( { 6
a_2^2 \over 35} - { f_2^2 \over 70} + { 5 u f_2^2 \over 14 X^2 } + { 2 u
a_2 \over 7 } + { 14 u^2 \over 5 } \right) y^6 + \cdots \right] \cr
a &= 1 + a_2 y^2 + \left( { 3 a_2^2 - f_2^2 \over 10 } \right) y^4 +
\left( { a_2^3 \over 10} - { 3 a_2 f_2^2 \over 35 } - { u f_2^2 \over 14
} \right) y^6 + \cdots \cr } \eqn\eq
$$ where $ u = \alpha X^2 /( 60 \pi ) $.
Numerical simulations indicate that neither $f_2 $ nor $ a_2 $ increase
with $X$ which, using \bci, \bcii, leads to $ f \sim 3 \pi x^2 / 2 X^2 $
and $ a \sim 1 - x^2 /X^2 $ for $ x \ll X $. Thus, as
$ X \rightarrow \infty $, $ f  \rightarrow 0 $ and $ a \rightarrow 1
$ for $x$ finite.

For $ x \lesim X $ the boundary conditions require $a
\simeq 0 $ and $f \gg 1 $; the first of the equations \eom\ can
then be approximated by
$$ f'' \simeq { \alpha \over 6 \pi } { f^3 \over X^2 } ; \qquad x
\lesim X \eqn\eq $$
whose solution (using \bci) reads
$$ f \simeq { 6 \pi X \over 4 + ( X - x ) \sqrt{ 3 \pi \alpha } } . \eqn\eq $$
Using these results we can evaluate the various thermodynamic quantities
for large $ X $. For example,
$$ \ecal - \ecal\lowti{ideal\ gas} \simeq \sqrt{{ 3 \pi \over 16 \alpha
}} \, { R^2 \over \tw^3 } ; \qquad
X \gg 1 \eqn\easympt $$ from which we find
$ | \ecal - \ecal\lowti{ideal\ gas} | / \ecal\lowti{ideal\ gas} \sim 1/X
$. All other thermodynamic
quantities exhibit this behavior: for large $T$
and fixed $R$  (corresponding to large
$X$) the system approaches a mixture of ideal gases.

We emphasize that this is not a
result of asymptotic freedom (when the running of the coupling is
included the large $X$ behavior will acquire logarithmic corrections),
but a property of the solutions to the differential equations. In the
infinite volume limit the charges are screened which requires $ \ao = 0
$ (see \defofdq).

It is also worth noticing that \easympt\ explicitly displays the
finite-volume corrections to the ideal gas results.

\chapter{Applications}

We now consider some applications of the above formalism. We first study
a system with vanishing baryon number (corresponding to $ \mu = 0 $), and
then consider the case of zero temperature.

\section{Zero baryon number}

\setbox2=\vbox to 170pt{\epsfysize=6 truein\epsfbox[0 -200 612 592]{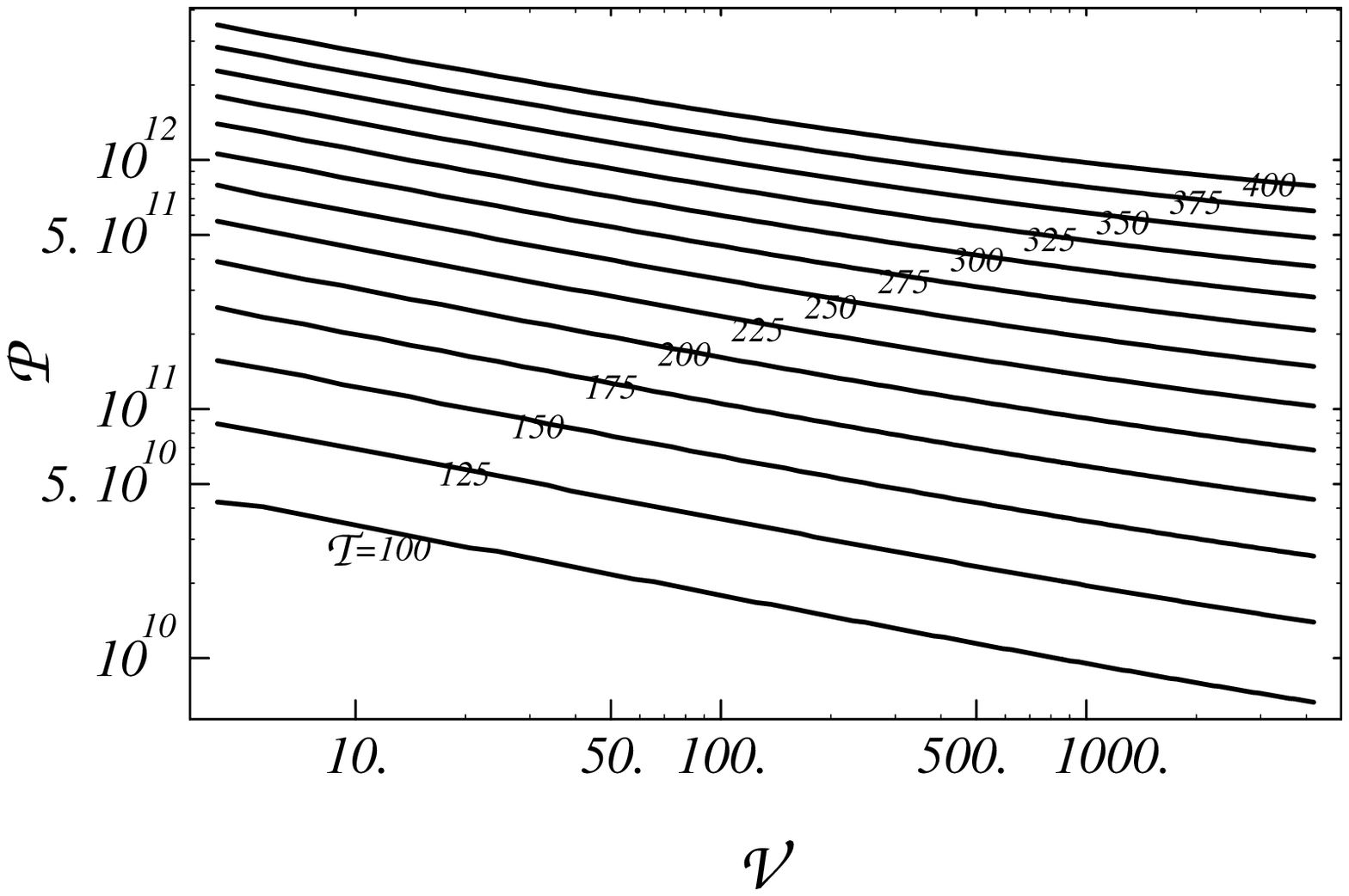}}
\setbox3=\vbox to 180pt{\epsfysize=5.8 truein\epsfbox[0 -180 612 612]{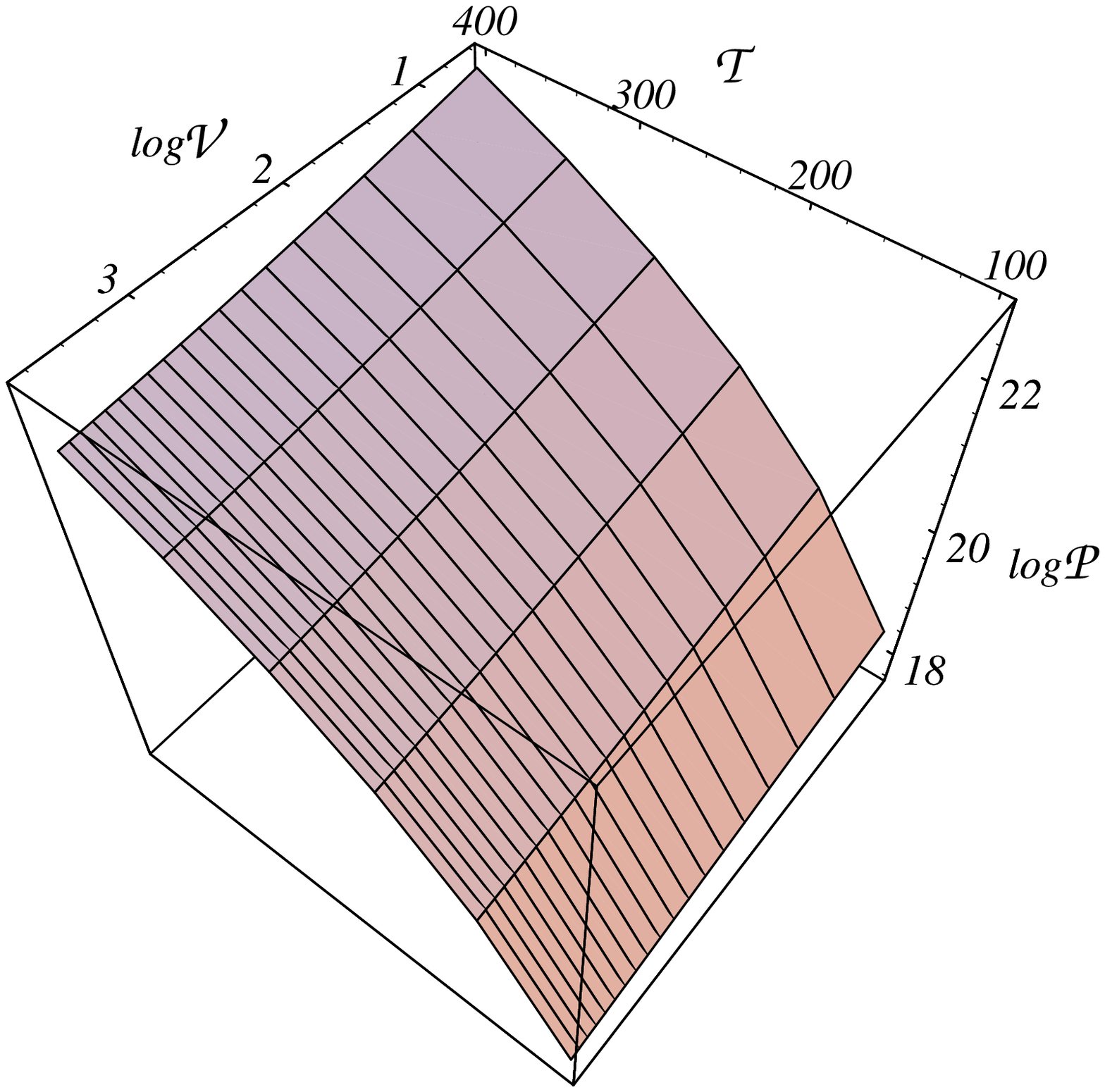}}

This situation is believed to be of relevance in relativistic heavy ion
collisions, such as those to be produced at RHIC~\refmark{\rhicref}, where,
in the standard picture, the nuclei will go through one another leaving
behind a region of hot quark-gluon plasma with zero baryon
number~\refmark{\zeroNatrhic}

The requirement  $ \ncal = 0 $ in \partnum\ corresponds to $
\mu = 0 $ which simplifies some of the expressions. In particular the
only scales in the system are the temperature and the volume.
The plot of the equation of state is given in
\FIG\myfiiia{Equation of state
within the \tf\ approximation for the $ \ncal = 0 $ case. The graph
displays the pressure as a function of the volume for several values of
the temperature \unitsPTV.}
\FIG\myfiiib{Three-dimensional rendition of the equation of state for
$ \ncal = 0 $ \unitsPTV}
Figs. \myfiiia\ and \myfiiib.

\centerline{\box2}
\thecaphyz{Equation of state
within the \tf\ approximation for the $ \ncal = 0 $ case. The graph
displays the pressure as a function of the volume for several values of
the temperature \unitsPTV.}

\centerline{\box3}
\thecaphyz{Three-dimensional rendition of the equation of state for $ \ncal = 0
$;
\unitsPTV}

\bigskip

We have determined $ \alpha $ by requiring
the solution to minimize the thermodynamic potential $ \Omega $ when the
energy density equals $ 4 \gev / \hbox{ fm }^3 $ at $ T = 150 \mev, \ R = 10
$fm (which is consistent with the expectations for RHIC); in this case
$ \alpha\simeq 1.568 $.

If we now allow the system to expand
adiabatically, we can use the above expressions to obtain
the relationship between $T$ and $R$  corresponding to
this process. This isentropic transformation describes (in an admittedly
oversimplified manner) the expansion of a quark-gluon plasma.
The entropy is gotten from \theentropy\ by setting $ \mu = 0 $,
the result is $$ \inv k \scal = { 13 \over
135 } X^3 + \inv3 \int_0^X dx \; f^2 (x ) . \eqn\eq $$ Since
$\scal $ is a function of $X$ only (a consequence of having only two scales
in the problem, $R$ and $ \tcal $), the equation for the
adiabats is $ X = $constant, or, equivalently $ \vcal \tcal^3 = $const.
corresponding to an adiabatic index $ \gamma = 4 $. Note that
the $ \int f^2 $ term in $ \scal $  modifies the usual  free
fermion gas relation $ S \propto \tcal ^3 $; the corrections are $ \sim
20\% $ (which is smaller than the corresponding contributions in the
case of small $X$, see section 2.5).

We can also easily determine the energy density for this isentropic
process. From the expression for the total energy in \energy\ it follows
that $ \ecal \tw $ is a function of $X$ only (for the $ \mu = 0 $
case). It follows that at constant entropy $ \ecal $ scales as $ \tcal
$. The energy density then will scale as $ \tcal / R^3 \propto \tcal^4
$, just as for an ideal gas of massless particles.

Using the expression for $ \scal $ we
obtain the heat capacity at constant volume, $$ \inv k C_V = X \deriva
\scal X {} . \eqn\CvfromS $$ In calculating this expression one must
remember that the boundary conditions to the TFQCD equations depend on
$X$, so that we should in fact write $ f = f( X ; x ) $; when
the partial derivative is taken in \CvfromS, $f$ must also be differentiated
under
the integral sign.

\section{Zero temperature}

\setbox2=\vbox to 220pt{\epsfysize=6 truein\epsfbox[0 -135 612 657]{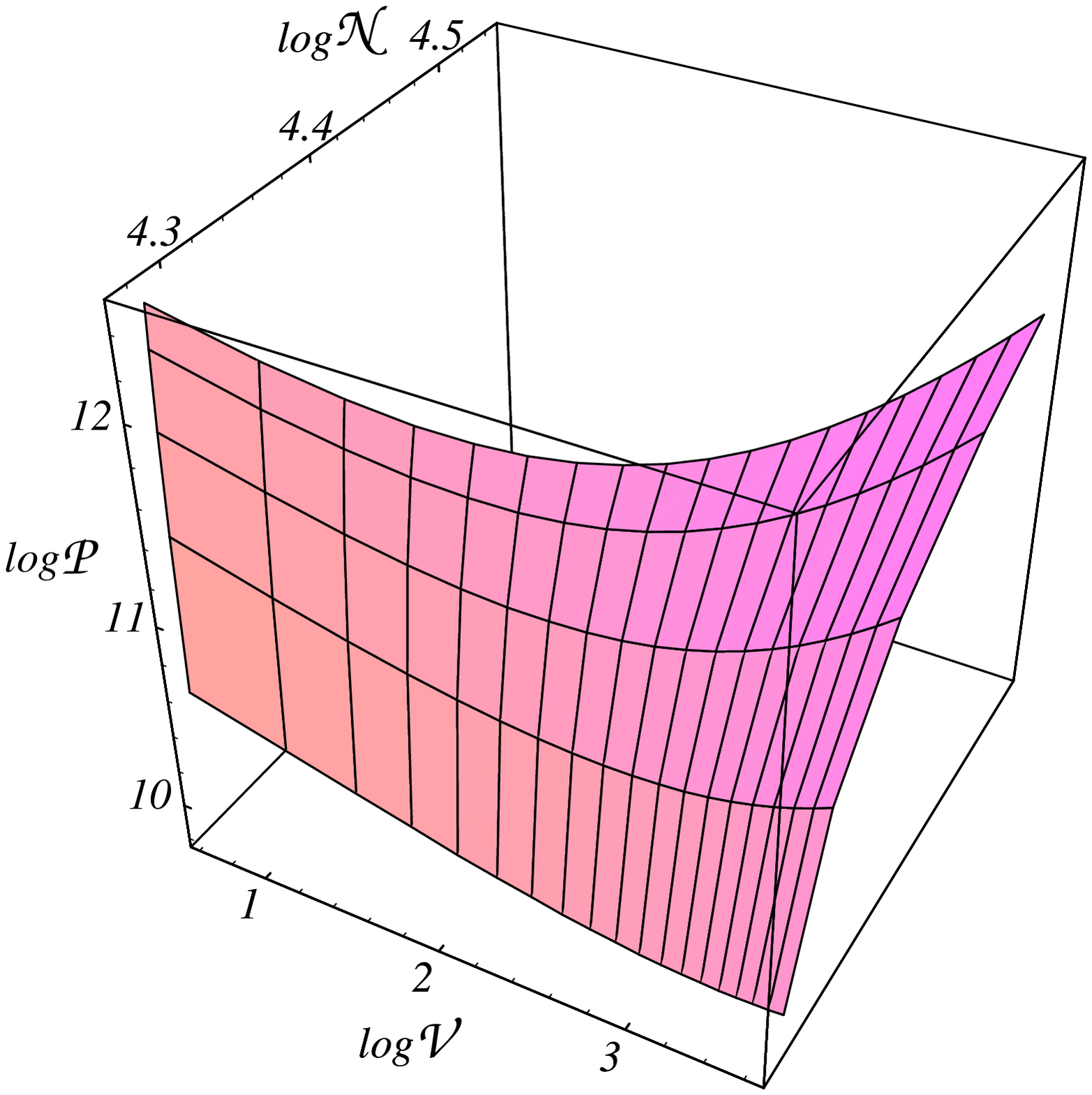}}
\setbox3=\vbox to 220pt{\epsfysize=6 truein\epsfbox[0 -135 612 657]{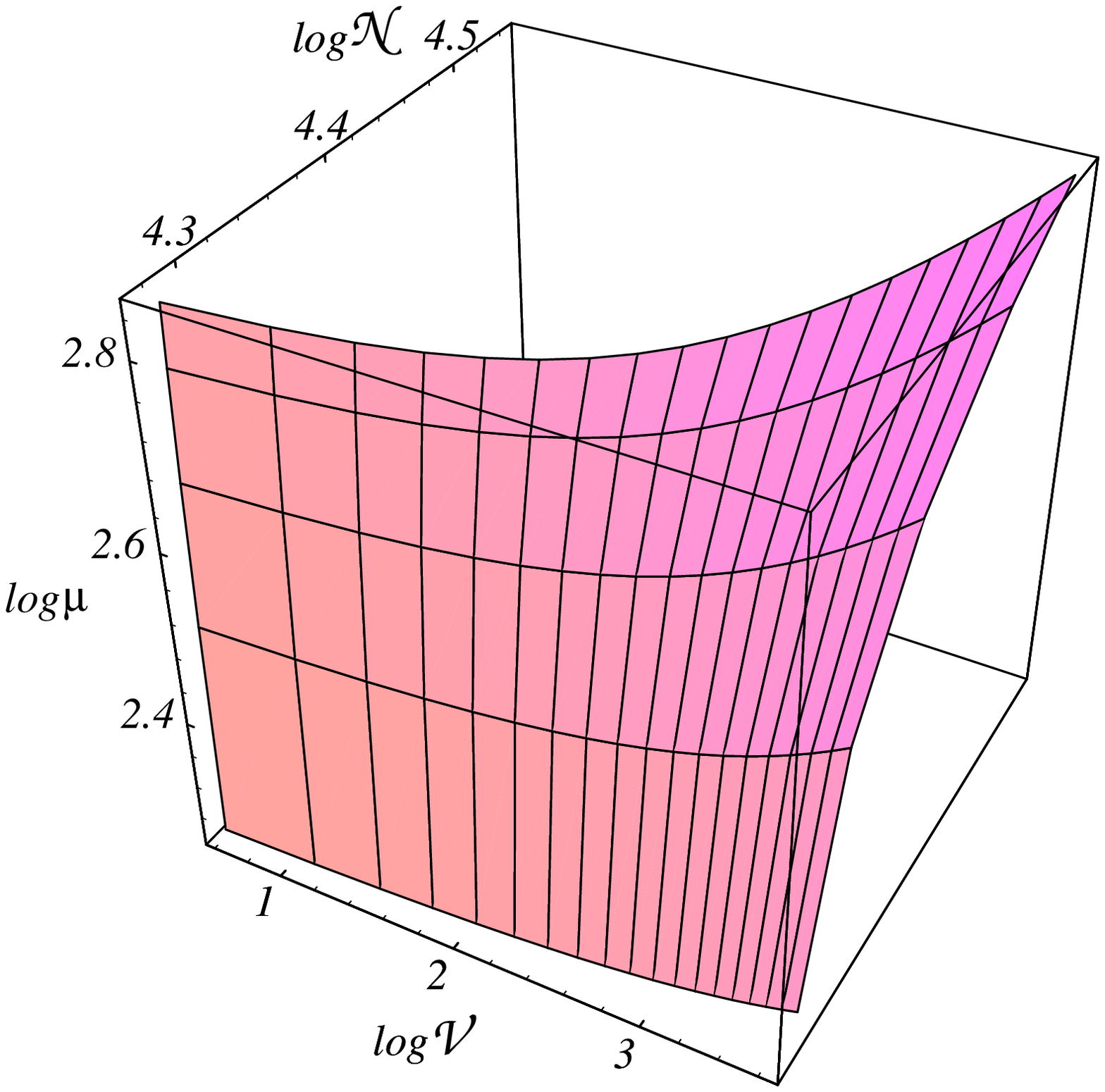}}
\setbox4=\vbox to 150pt{\epsfysize=6 truein\epsfbox[0 -200 612 592]{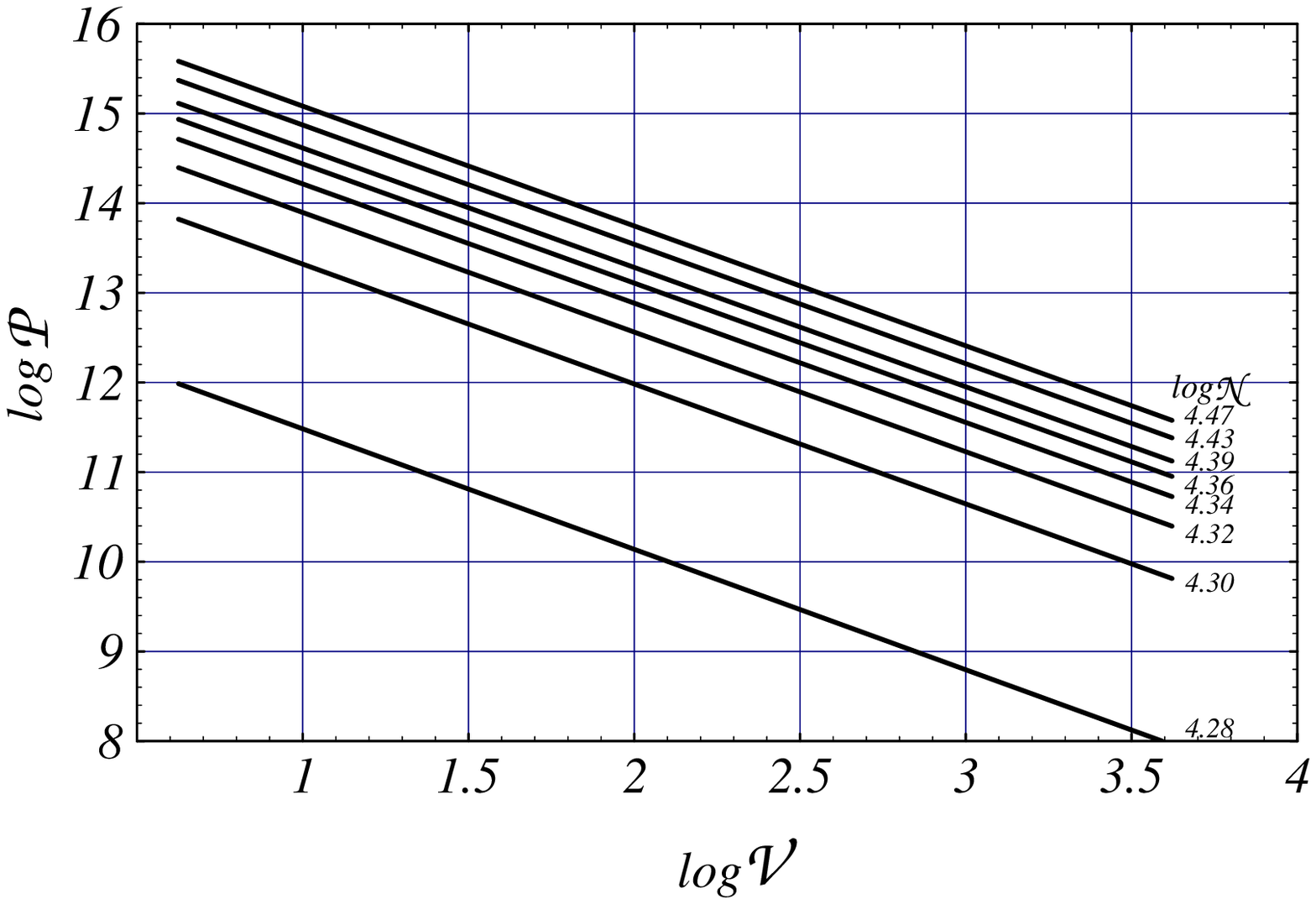}}

We now turn to the case of zero temperature; the dimensional quantities
in the system are now $ \mu $ and $R$. In this case all dimensionless
quantities such as $ \ecal/\mu $ will be functions of $ R \mu $ only. The
chemical potential is determined in terms of $R$ and $ \ncal $ using
\partnum\ but this must be done numerically since the
non-ideal gas term is significant and cannot be ignored. The
plot of the equation of state for this case is given in
\FIG\myfiva{The equation of state within the \tf\ approximation for the
case zero temperature case; \unitsPVN}
Fig. \myfiva.

\centerline{\box2}
\thecaphyz{The equation of state within the \tf\ approximation for the
case zero temperature case; \unitsPVN}

The equivalent contour plot for various values of $N$ is presented in
\FIG\myfivc{The equation of state within the \tf\ approximation for the
case zero temperature case; \unitsPVN}
Fig. \myfivc

\centerline{\box4}
\thecaphyz{Pressure as a function of volume at zero temperature, for
various values of $ \log \ncal $; \unitsPVN }

The equation of state (for the range of variables presented in figure
\myfivc) is well represented by the equation $$ \pcal \vcal^{4/3} =
z(N) ; \qquad \log z(N) \simeq 12.82+ 5.46 \left[ \log \left( { \ncal \over
1.91 \times 10^4 } \right) \right]^{1/4} \eqn\eq $$ The $\pcal \vcal^{4/3} $
behavior is a result of simple scaling arguments and is therefore
present here as well as for small $X$. In contrast, the  $ \ncal $
dependence of the equation of state is radically different (cf. \eoslowX).

We have also determined the chemical potential as a function of
temperature and volume. The result is presented in
\FIG\myfivb{The chemical potential as a function of volume and baryon
number at zero temperature; \unitsPVmu}
Fig. \myfivb.

\centerline{\box3}
\thecaphyz{The chemical potential as a function of volume and baryon
number at zero temperature; \unitsPVmu}

 As $ \tcal \rightarrow 0 $ the entropy goes to zero linearly,
$$ { \scal \over
\pkt } \;{ \buildrel \tcal \rightarrow 0  \over \longrightarrow } \;
{ k \over \sqrt{3} \, \mu } \left[ \left( { X_o \over 3 } \right)^3
+ \inv3 \int_0^{X_o} dx \, f^2 ( x ) \right] ,
\quad X_o = 2 \sqrt{3} \,R \mu , \eqn\eq $$ since the fermionic
contribution dominates in this limit; we then also have
$ C_V = \scal $.

\chapter{Extensions of the method}

The inclusion of more flavors is quite straightforward, the charges
generated by each simply add. Possible computational  difficulties arise
when the fermion mass cannot be neglected (as is the case for the strange
quark) for in this case a closed form for the fermionic partition
function is not available. We will not pursue here this situation further as
it involves no new concepts.

A more interesting extension is obtained by considering $ \su3 $ as the
gauge group. In this case there are two important modifications. First,
within each subvolume $ \dv $, though we still have $ \bar A_0 = $constant
and diagonal, this now implies $ \bar A_0 = g ( \ao \lambda_3 +
\bo \lambda_8 )/( 2 i ) $. In general $ \bo \not= 0 $, so in this case
we will have additional contributions depending on this new potential.
The TFQCD equations are derived in the same way as for the $ \su2 $
case.  Therefore the presence of the gauge field
is summarized by the replacements $$ \eqalign{ \mu \rightarrow
& \mu +  { g \over 2 } \left( \ao + \bo / \sqrt{3} \right) , \cr
& \mu +  { g \over 2 } \left( - \ao + \bo / \sqrt{3} \right) , \cr
& \mu - g \bo / \sqrt{3} , \cr } \eqn\eq $$ in $ Z_0 $ (Eq. \freeZ)
The resulting fermionic partition function is then $$ \eqalign{
\ln Z_\psi =&  {
\beta \dv \over 4 \pi^2 } \Biggl[ { g^4 \over 24} \left( \ao^2 + \bo
{}^2 \right)^2 + { \mu g^3 \over 3 \sqrt{ 3 } } \bo \left( 3 \ao^2 -
\bo{}^2 \right) \cr & \quad
+ { g^2 \over 12 \tw^2 } \left( \ao^2 + \bo {}^2 \right)
+ \mu^4 + 2 \mu^2 \left( \pkt \right)^2 + { 7 \over 15 } \left( \pkt
\right)^4 \Biggr] .\cr } \eqn\eq $$

Using this result we obtain the charge densities $$ \eqalign{ \qden_3  =
& { g \ao \over 6 \pi^2 } \left[ { g^2 \over 4 } \left( \ao^2 +
\bo{}^2 \right) + \sqrt{3} \; g \mu \bo  + \inv{ 4 \tw^2 }
\right] , \cr \qden_8 =
& { g \over 2 \sqrt{3} \, \pi^2 } \left[ { g^2 \over 4 } \bo \left(
\ao^2 + \bo{}^2 \right) + { \sqrt{ 3 } \over 2 } g \mu \left(
\ao^2 - \bo{}^2 \right) + { \bo \over 4 \tw^2 } \right] . \cr } \eqn\eq
$$

The second modification concerns the form of
the spherically symmetric Ansatz for the background gauge potentials.
For $ \su3$ a possible Ansatz takes the form (now including a
contribution in the $ \lambda_8$ direction)
$$ \eqalign{
\bar A_a^0 &= \ao\rh_a ; \quad ( a = 1 , 2 , 3 ) \cr
\bar A_a^i &= \epsilon_{ i a j } \rh_j \left( { 1 + \varphi_2
\over r } \right) +  \left( \delta_{ i a } - \rh_i \rh_a \right) {
\varphi_1 \over r } + \rh_i \rh_a \ai ; \quad ( a = 1 , 2 , 3 ) \cr
\bar A_8^0 &= \bo . } \eqn\suiiiAnsatz $$
Note however that the choice of the $ \su2 $ subgroup in which the
potentials $ \bar A_a^\mu , \ ( a = 1, 2, 3 )$ reside is arbitrary,
and that it
costs no energy to change from one such subgroup to another; these
degrees of freedom are included through a set of
collective coordinates~\refmark{\collcoords}.
The full Ansatz we use is then (we define $
\bar A^\mu = \lambda_n \bar A_n^\mu / ( 2 i ) $ where the $ \lambda_n$ denote
the
usual Gell-Mann matrices) $$ \bar A^\mu \rightarrow \bar A^\mu =
U^\dagger \bar A^\mu U \eqn\collcoordAnsatz $$ where the $ \bar A_n^\mu $
are given in \suiiiAnsatz\ and $U$ is a time dependent $ \su3 $ matrix.

The Lagrangian for the background gauge fields
then becomes $$ \half \tr \bar F_{ \mu \nu } ^2
\rightarrow \half \tr \bar F_{ \mu \nu } ^2 + 2 \tr F^{ 0 i } \left[ \bar A_i ,
\rBB \right] + \tr\left\{ \left[ \bar A_i , \rBB \right] \left[ \bar A^i , \rBB
\right]
\right\} \eqn\Fsqbar $$ where $$ \rBB = \dot U \, U^\dagger , \eqn\eq $$
and $ \bar F $ is the field strength corresponding to $ \bar A $.
When the form of the gauge potentials in the $\su2$ subgroup takes
the form \suiiiAnsatz, $ \rBB $ should have no
components along the generators of the $ \su2 $ subgroup generated by $
\lambda_{ 1 , 2 , 3} $, that is we
take $$ \rBB = \sum_{ n = 4 }^8 \inv{2 i } \lambda_n \rBB^n \eqn\eq $$
which considerably simplifies \Fsqbar.
The corresponding action is~\foot{The simplicity of this
result is a consequence of the fact that $U$ is made to
reside in $ \su3 / \su2 \sim S^5 $, a five-dimensional sphere, where the
number of invariants is very limited. The solutions of to the classical
equations of motion for $U$ are geodesics representing a motion along the great
circles of $ S^5 $.} $$ S = \int d^4 x \inv{ 2 g^2 }
\tr F_{ \mu \nu }^2  + \half c^2 \int dt \; \tr \dot U^\dagger \dot U ;
\quad c^2 = \inv \alpha \int_0^R dr \; ( a - 1 )^2 . \eqn\eq $$
Numerically the coefficient $c$ can be very large (for the numerical
solutions presented $ c \sim 3.5 \times 10^3 \tw $)

We will use the notation $$ \bo= { h ( r ) \over r g } . \eqn\eq $$
whence the TFQCD equations become (a prime denotes a derivative with
respect to $ x = r / \tw $), $$ \eqalign{ f'' &= { 2 a^2 \over x^2 } f +
{ \alpha \over 6 \pi } f \left[ { f^2 + h^2 \over x^2 } + \left( 4
\sqrt{3} \; \tw \mu \right) { h \over x } + 1 \right] \cr a'' &= { a^2 -
f^2 - 1 \over x^2 } a  \cr h'' &= { \alpha \over 6 \pi } \left[
{ h ( f^2 + h^2 ) \over x^2 } + \left( 2 \sqrt{3} \; \tw \mu \right)
{ f^2 - h^2 \over x } + h \right] \cr } \eqn\eq $$ which can be solved
using the same methods as before.  Note that $ h = 0 $ is not allowed
when $ \mu \not= 0 $.

For the interesting case $ \mu = 0 $, $ h = 0 $ is a solution
to the above equations.
Hence, for zero baryon number, the previous solutions also satisfy the $
\su3 $ TFQCD equations. It does not follow, however, that these solutions
again minimize the thermodynamic potential. Note also that even in the
case $ h = 0 $ there is an additional contribution to the thermodynamic
functions from the collective variables $U$

We will not pursue this case further in this paper. A realistic
investigation of the $ \su3 $ case requires we include mass
term for the (strange) quarks, and also the contributions of the
collective coordinates to the thermodynamics of the system. We will
consider these issues in a forthcoming publication.

\chapter{Conclusions}

We have presented an approximate treatment of QCD based on the same
ideas as the Thomas Fermi atom. Within this framework the
thermodynamics of the system can be derived and the results can be
compared with the experimental results which will soon be available.

The method is based on a subdivision of the system into subvolumes which
are still large enough to be considered statistical systems. These
subvolumes interact through an average gauge field whose sources
are the thermodynamically averaged non-Abelian
charges for the subvolumes. These charges, though small, are not completely
screened due to the assumed smallness of the subvolumes.

The formalism was developed in this paper for the simplified case of an
$\su2$ gauge group, though we did provide a brief discussion of the
modifications required for and $ \su3 $ theory. We also ignored fermion
masses and all interactions between the partons
inside each of the subvolumes. Nonetheless we found that the numerical
values for the pressure in the small $ \ncal $ case are in rough
agreement with the bag-model calculations.

For large temperatures, or densities ($X \gg1$) the solutions to the equations
of
motion are such that all thermodynamic quantities approach those of a
mixture of ideal gases, with $ 1/X $ measuring the deviation from this
limiting behavior. This feature is not related to asymptotic freedom but
a result of screening.

In the limit $ R \rightarrow \infty $ we have
$ f =0 $ and $ a = 1 $, and the equation of state reduces to
that of an ideal gas. This model then provides an approximation to the
finite-volume corrections to the ideal gas, this is explicitly
demonstrated in \easympt\ which gives the surface corrections to the
energy of the system.

A realistic calculations must be
performed for an $ \su3 $ gauge theory with massive fermions; the
partition function inside each subvolume should be evaluated to the
highest order available (or possible) in perturbation theory.
The inclusion of radiative corrections will induce, among other things, a
dependence of the (now running) coupling constants on the temperature
and chemical potential. For the present calculation no such effects were
included. Finally one should also include finite volume effects as well
as the corrections induced by the gluonic partition function. We will
investigate such realistic situations in a forthcoming publication.

We found two solutions to the equations of motion satisfying the
boundary conditions and which minimize the thermodynamic potential $ \Omega
$. Both lead to the same thermodynamics and appear indistinguishable
except near the origin (at least within numerical errors). A complete
study of the behavior of these solutions under non-spherical perturbations
along the lines of Ref~\jr\
is required to determine the one which is most stable. We have
not performed such an investigation since the presence of two such
solutions does not alter the thermodynamics derived within the TFQCD
approach.

The above treatment was not based on a semi-classical expansion of the
partition function for the complete system. It is indeed possible to
consider such an approach and use \defofzpsi\ as an approximation to the
fermionic contribution. Then the integration over the gauge fields can
be approximated by a saddle point method. We have not done this because
the effective action which is to be minimized in the last step is, due
to the \tf\ approximations used to obtain $ Z_\psi $, unbounded from
below. It is found that the
solutions will soon violate the \tf\ conditions and the method is not
consistent; this is displayed explicitly in the appendix for the case
of QED. In contrast, the approach described in the above is consistent
with the original approximations.

\appendix

\def\sef{{ S \lowti{eff}}}

In this appendix we present a semi-classical calculation of the
partition function of QED using the \tf\ approximation for the fermionic
partition function. The general expression is
$$ Z = \int [dA] [d \psi ] [d\bar\psi] e^S \eqn\eq $$ where $ S = S_g +
S_\psi $, the first term denoting the gauge contribution, the second all
terms involving the fermions. By definition we have
$$ Z_\psi = \int [d \psi ] [d\bar\psi] e^{ S_\psi } \eqn\eq $$
which is approximated by $ Z_\psi \simeq \int d^4 x \pcal_\psi $, where
$$ \pcal_\psi = \inv{ 12 \pi^2 } \left[ ( \mu + e \phi )^4 + 2 \left( \pkt
\right)^2 ( \mu + e \phi )^2 + { 7 \over 15 }\left( \pkt \right)^4
\right] \eqn\eq $$ where $ \phi $ denotes the electrostatic potential,
and $e$ the charge of the fermions (only one flavor is considered).
Note that $ \pcal_\psi $ is positive definite.

Assuming spherical symmetry the gauge potentials are of the form, $
\phi = \phi(r) , \ \AA = a ( r ) \rh $. Choosing the $ a = 0 $ gauge
gives the following expression for the partition function
$ Z = \int [d \phi ] \exp \sef $, where $$ \sef = 4
\pi \int_0^\beta dt \; \int_0^R dr \; r^2 \left[ - \half \left( \phi'
\right)^2 + \pcal_\psi \right] \eqn\eq $$ where $ \beta $ denotes the
inverse temperature, $R$ is the radius of the spherical vessel
containing the system, a dash denotes a derivative with respect to $r$,
the radial coordinate, and $t$ denotes the Euclidean time variable.

The integrand in $ \sef $ is not positive definite. Consider for example
$ \phi = \phi_0 \cos ( k r + \nu ) $ for constant $ \nu $. If $ \phi_0 $
is sufficiently small and $k$ sufficiently large, the first term in $
\sef $ will dominate; the larger $k$, the more negative $ \sef $
becomes. The problem in this case is that these expressions for the
scalar potential violate the \tf\ condition which requires $ \tw
\phi' / \phi \ll 1 $. this shows that a semi-classical treatment of the
partition function is inconsistent with the \tf\ approximation. We have
verified that the same problems arise in the non-Abelian case.

\ack

This work was supported in part through funds provided by DoE through
grant DE-FG03-92ER40837. We are grateful to L. Brown and L. Jacobs
for several interesting comments. J.W. and
P.K. also acknowledge the hospitality of the 1995 Aspen Center for Physics
where part of this work was completed.

\advance \baselineskip by -1 pt
\refout
\figout

\bye